\documentclass[10pt, a4paper, fleqn]{article}
\usepackage{titling}
\title{The Collisional Particle-In-Cell Method for the Vlasov-Maxwell-Landau Equations}

\newcommand{\authorPDF}{Bailo, Carrillo, Hu.}
\newcommand{\subjectPDF}{35Q83; 35Q61; 35Q84; 65M75; 76M28. }
\newcommand{\keywordsPDF}{Collisional plasma;
	Vlasov-Maxwell-Landau equations;
	particle-in-cell (PIC) methods.
}
 \usepackage{authblk}

\author[1]{Rafael Bailo}
\author[1]{Jos\'{e} A. Carrillo}
\author[2]{Jingwei Hu}

\affil[1]{
	Mathematical Institute, University of Oxford
}
\affil[ ]{
	OX2 6GG Oxford, United Kingdom
}
\affil[ ]{\textit{bailo@maths.ox.ac.uk, carrillo@maths.ox.ac.uk}}

\affil[ ]{}

\affil[2]{
	Department of Applied Mathematics, University of Washington
}
\affil[ ]{
	Seattle, WA 98195, United States of America
}
\affil[ ]{\textit{hujw@uw.edu}} 
\usepackage[margin=1.1in]{geometry}

\makeatletter
\let\newtitle\@title
\let\newauthor\@author
\let\newdate\@date
\makeatother
\usepackage{fancyhdr}
\usepackage{setspace}
\usepackage[
	hypertexnames=false,
	pdftex,
	pdftitle={\thetitle. \authorPDF},
	pdfauthor={\authorPDF},
	pdfsubject={\subjectPDF},
	pdfkeywords={\keywordsPDF},
]{hyperref}
\hypersetup{
	colorlinks=true,
	linkcolor=blue,
	citecolor=blue,
	urlcolor=blue,
	linktocpage
}

\usepackage[utf8]{inputenc}
\usepackage[T1]{fontenc}
\usepackage{amsmath}
\usepackage{amsthm}
\usepackage{amsfonts}
\usepackage{amssymb}
\usepackage{graphicx}
\usepackage{mathtools}
\usepackage{ifthen}
\usepackage{dsfont}

\usepackage[UKenglish]{babel}

\usepackage[dvipsnames]{xcolor}

\definecolor{color1}{RGB}{0, 121, 178}
\definecolor{color2}{RGB}{255, 124, 37}
\definecolor{color3}{RGB}{37, 160, 55}
\definecolor{color4}{RGB}{220, 32, 44}
\definecolor{color5}{RGB}{147, 104, 186}
\definecolor{color6}{RGB}{143, 85, 76}
\definecolor{color7}{RGB}{230, 119, 192}
\definecolor{color8}{RGB}{127, 127, 127}
\definecolor{color9}{RGB}{192, 188, 55}
\definecolor{color10}{RGB}{0, 191, 206} 

\newcounter{review}

\newcommand{\ntcreview}[3]{\refstepcounter{review}{\color{#2}{\textbf{[#1]}: #3}}}

\newcommand{\creview}[3]{\ntcreview{#1}{#2}{#3}
	\addcontentsline{tor}{subsection}{\thereview~\textbf{[#1]}:~#3
	}}

\newcommand{\review}[2]{\creview{#1}{blue}{#2}}

\makeatletter

\newcommand\listreviewname{List of Reviews}
\newcommand\listofreviews{\section*{\listreviewname}\@starttoc{tor}}
\makeatother

\usepackage[all]{nowidow}

\newcommand{\subjectclassification}[1]{

	{\small\textbf{\textit{AMS Subject Classification --- }} #1}

}
\newcommand{\keywords}[1]{

	{\small\textbf{\textit{Keywords --- }} #1}

}

\renewcommand\lll\MoveEqLeft

\usepackage{tikz}
\usetikzlibrary{patterns}
\usetikzlibrary{decorations.pathreplacing}
\usetikzlibrary{calc}
\usetikzlibrary{fadings}
\usetikzlibrary{shapes}
\usetikzlibrary{plotmarks}
\usetikzlibrary{arrows, decorations.markings}
\tikzset{thicker line small arrows m/.style args={#1in#2}{
			draw=#2,
			solid,
			line width=#1,
			shorten >=1mm,
			decoration={
					markings,
					mark=at position 1.0 with {\arrow[fill=#2,thin]{triangle 90}}
				},
			postaction={decorate}
		}}

\usepackage{pgfplots}
\pgfplotsset{compat=1.16}

\usepackage{float}
\usepackage{subcaption}
\usepackage[section]{placeins}
\usepackage{rotating}

\usepackage{array}
\newcolumntype{L}[1]{>{\raggedright\let\newline\\\arraybackslash\hspace{0pt}}m{#1}}
\newcolumntype{C}[1]{>{\centering\let\newline\\\arraybackslash\hspace{0pt}}m{#1}}
\newcolumntype{R}[1]{>{\raggedleft\let\newline\\\arraybackslash\hspace{0pt}}m{#1}}
\usepackage{booktabs}
\usepackage{tabularx}
\usepackage{multirow}

\usepackage{titling}

\usepackage[algoruled]{algorithm2e}

\usepackage[capitalise]{cleveref}

 \usepackage{accents}

\newcommand\term\emph

\numberwithin{equation}{section}

\usepackage{autobreak}

\usepackage{enumerate}

\usepackage{apptools}

\makeatletter
\def\@maketitle{\newpage
	\begin{center}\let \footnote \thanks
		{\LARGE\bfseries \@title \par}\vskip 2.5em{\large
				\lineskip .5em\begin{tabular}[t]{c}\@author
				\end{tabular}\par}\vskip 1em{\large \@date}\end{center}\par
	\vskip 1.5em}
\makeatother



\usepackage{amsthm}
\theoremstyle{plain}
\newtheorem{theorem}{Theorem}[section]

\theoremstyle{remark}
\newtheorem{remark}[theorem]{\bf Remark}

\usepackage{esint}

\def\XXint#1#2#3{{\setbox0=\hbox{$#1{#2#3}{\int}$ }
			\vcenter{\hbox{$#2#3$ }}\kern-.6\wd0}}

\renewcommand{\th}{\textsuperscript{th} }

\DeclarePairedDelimiter{\prt}{(}{)}
\DeclarePairedDelimiter{\brk}{[}{]}

\DeclarePairedDelimiter{\abs}{|}{|}
\DeclarePairedDelimiter{\norm}{\|}{\|}
\DeclarePairedDelimiter{\set}{\{}{\}}

\DeclarePairedDelimiter{\inn}{\langle}{\rangle}

\usepackage{suffix}

\newcommand{\inner}[2]{\inn{#1,#2}}
\WithSuffix\newcommand\inner*[2]{\inn*{#1,#2}}

\DeclarePairedDelimiter{\positive}{(}{)^{+}}
\DeclarePairedDelimiter{\negative}{(}{)^{-}}

\newcommand\pos\positive
\renewcommand\neg\negative

\WithSuffix\newcommand\pos*{\positive*}
\WithSuffix\newcommand\neg*{\negative*}

\newcommand{\R}{{\mathbb{R}}}

\newcommand{\dimx}{{d_x}}
\newcommand{\dimv}{{d_v}}

\renewcommand{\L}[1]{{L^{#1}}}

\newcommand{\Ltwo}{\L{2}}

\newcommand{\pnorm}[2]{\norm{#2}_{\L{#1}}}
\WithSuffix\newcommand\pnorm*[2]{\norm*{#2}_{\L{#1}}}

\newcommand{\psnorm}[3]{\norm{#3}_{\L{#1}(#2)}}
\WithSuffix\newcommand\psnorm*[3]{\norm*{#3}_{\L{#1}(#2)}}

\newcommand{\pnormp}[2]{\pnorm{#1}{#2}^{#1}}
\WithSuffix\newcommand\pnormp*[2]{\pnorm*{#1}{#2}^{#1}}

\newcommand{\psnormp}[3]{\psnorm{#1}{#2}{#3}^{#1}}
\WithSuffix\newcommand\psnormp*[3]{\psnorm*{#1}{#2}{#3}^{#1}}

\newcommand\svec\vec

\renewcommand{\vec}{\mathbf}
\renewcommand{\svec}{\boldsymbol}

\newcommand{\conv}{\ast}

\renewcommand{\d}{\mathrm{d}}
\newcommand{\dd}{\mathop{}\!\d}

\newcommand{\der}[2]{\frac{\d #1}{\d #2}}

\newcommand{\vder}[2]{\frac{\delta #1}{\delta #2}}

\newcommand{\dx}{\dd x}

\newcommand{\grad}{\nabla}
\newcommand{\gradx}{\nabla_x}

\newcommand{\gradv}{\nabla_v}

\renewcommand{\div}{\nabla\cdot}
\newcommand{\divx}{\gradx\cdot}

\newcommand{\divv}{\gradv\cdot}

\newcommand{\curlx}{\gradx\times}

\newcommand{\pt}{\partial_t}
\newcommand{\px}{\partial_x}

\newcommand{\Dt}{\Delta t}
\newcommand{\Dx}{\Delta x}

\newcommand{\h}{_{h}}

\newcommand{\nhalf}{1/2}

\renewcommand{\i}{_{i}}

\newcommand{\ih}{_{i+\nhalf}}

\newcommand{\n}{^{n}}
\newcommand{\np}{^{n+1}}

\newcommand{\ppr}{(r)}

\newcommand{\Wr}{^{W,\,\ppr}}

\newlength{\dhatheight}

\renewcommand{\S}{^{S}}

\ifthenelse{\isundefined{\Wr}}{
	\newcommand{\Wr}{^{W,\,\ppr}}
}{
	\renewcommand{\Wr}{^{W,\,\ppr}}
}

\newcommand{\Np}{N}
\newcommand{\Nc}{N_c}
\newcommand{\Nx}{N_{x}}

\newcommand{\Nv}{N_{v}}
\newcommand{\Nvx}{N_{v_x}}
\newcommand{\Nvy}{N_{v_y}}

\newcommand{\fN}{f^\Np}
\newcommand{\ftN}{\tilde{f}^\Np}
\newcommand{\U}{\mathcal{U}}

\newcommand{\ex}{\eta}
\newcommand{\ev}{\varepsilon}
\newcommand{\Sx}{\psi_\ex}
\newcommand{\Shx}{\hat{\psi}_\ex}
\newcommand{\Sxh}{\hat{\psi}_h}
\newcommand{\Sv}{\varphi_\ev}
\newcommand{\Shv}{\hat{\varphi}_\ev}

\newcommand{\rxv}{_{\ex,\,\ev}}

\newcommand{\convxv}{\conv_{x,v}}

\newcommand{\xst}{x_*}
\newcommand{\vst}{v_*}
\newcommand{\dv}{\dd v}
\newcommand{\dxst}{\dd \xst}
\newcommand{\dvst}{\dd \vst}

\renewcommand{\H}{\mathcal{H}}

\renewcommand{\S}{\mathcal{S}}

\newcommand{\gradvst}{\nabla_{\vst}}

\newcommand{\ion}{_{\textrm{ion}}}

 \usepackage{todonotes}

\DeclareCaptionFormat{cont}{#1 (cont.)#2#3\par}

\newif\ifskiptable

\pgfplotsset{colormap={hsv}{
			hsb(0.00cm)=(0.00,0,0.95);
			hsb(0.05cm)=(0.05,1,1);
			hsb(0.10cm)=(0.10,1,1);
			hsb(0.15cm)=(0.15,1,1);
			hsb(0.20cm)=(0.20,1,1);
			hsb(0.25cm)=(0.25,1,1);
			hsb(0.30cm)=(0.30,1,1);
			hsb(0.35cm)=(0.35,1,1);
			hsb(0.40cm)=(0.40,1,1);
			hsb(0.45cm)=(0.45,1,1);
			hsb(0.50cm)=(0.50,1,1);
			hsb(0.55cm)=(0.55,1,1);
			hsb(0.60cm)=(0.60,1,1);
			hsb(0.65cm)=(0.65,1,1);
			hsb(0.70cm)=(0.70,1,1);
			hsb(0.75cm)=(0.75,1,1);
			hsb(0.80cm)=(0.80,1,1);
			hsb(0.85cm)=(0.85,1,1);
			hsb(0.90cm)=(0.90,1,1);
			hsb(0.95cm)=(0.95,1,1);
			hsb(1.00cm)=(1.00,1,1);
		}
}

\pgfplotsset{colormap={hsvSoft}{
			hsb(0.00cm)=(0.00,0,0.95);
			hsb(0.05cm)=(0.05,1,1);
			hsb(0.10cm)=(0.10,1,1);
			hsb(0.15cm)=(0.15,1,1);
			hsb(0.20cm)=(0.20,1,1);
			hsb(0.25cm)=(0.25,1,1);
			hsb(0.30cm)=(0.30,1,1);
			hsb(0.35cm)=(0.35,1,1);
			hsb(0.40cm)=(0.40,1,1);
			hsb(0.45cm)=(0.45,1,1);
			hsb(0.50cm)=(0.50,1,1);
			hsb(0.55cm)=(0.55,1,1);
			hsb(0.60cm)=(0.60,1,1);
			hsb(0.65cm)=(0.65,1,1);
			hsb(0.70cm)=(0.70,1,1);
			hsb(0.75cm)=(0.75,1,1);
			hsb(0.80cm)=(0.80,1,1);
			hsb(0.85cm)=(0.85,1,1);
			hsb(0.90cm)=(0.90,1,1);
			hsb(0.95cm)=(0.95,1,1);
			hsb(1.00cm)=(0.00,0,0.95);
		}
}

\pgfplotsset{colormap={viridisSoft}{
			rgb255=(242, 242, 242);
			rgb255=(242, 242, 242);
			rgb=(0.28026,0.1657,0.4765);
			rgb=(0.26366,0.23763,0.51877);
			rgb=(0.23744,0.3052,0.54192);
			rgb=(0.20862,0.36775,0.55267);
			rgb=(0.18225,0.42618,0.55711);
			rgb=(0.1592,0.48224,0.55807);
			rgb=(0.13777,0.53749,0.5549);
			rgb=(0.12115,0.59274,0.54465);
			rgb=(0.12808,0.64775,0.5235);
			rgb=(0.18065,0.7014,0.48819);
			rgb=(0.27415,0.75198,0.4366);
			rgb=(0.39517,0.79747,0.36775);
			rgb=(0.53561,0.83578,0.2819);
			rgb=(0.68895,0.86545,0.18272);
			rgb=(0.84557,0.88733,0.0997);
			rgb=(0.99324,0.90616,0.14394)
		}
}

\pgfplotsset{/pgfplots/colormap={bone}{[1cm]rgb255(0cm)=(0,0,0) rgb255(3cm)=(84,84,116) rgb255(6cm)=(167,199,199) rgb255(8cm)=(255,255,255)}
}

\pgfplotsset{colormap={jet}{
			rgb=(0.0,0.0,0.498);
			rgb=(0.0,0.0,1.0);
			rgb=(0.0,0.498,1.0);
			rgb=(0.0,1.0,1.0);
			rgb=(0.498,1.0,0.498);
			rgb=(1.0,1.0,0.0);
			rgb=(1.0,0.498,0.0);
			rgb=(1.0,0.0,0.0);
			rgb=(0.498,0.0,0.0)
		}
}

\pgfplotsset{colormap={spectral}{
			rgb=(0.0,0.0,0.0);
			rgb=(0.0933,0.0,0.1067);
			rgb=(0.1867,0.0,0.2133);
			rgb=(0.28,0.0,0.32);
			rgb=(0.3734,0.0,0.4266);
			rgb=(0.4667,0.0,0.5333);
			rgb=(0.48,0.0,0.5466);
			rgb=(0.4933,0.0,0.56);
			rgb=(0.5067,0.0,0.5733);
			rgb=(0.52,0.0,0.5867);
			rgb=(0.5333,0.0,0.6);
			rgb=(0.4266,0.0,0.6133);
			rgb=(0.32,0.0,0.6267);
			rgb=(0.2133,0.0,0.64);
			rgb=(0.1067,0.0,0.6534);
			rgb=(0.0,0.0,0.6667);
			rgb=(0.0,0.0,0.7067);
			rgb=(0.0,0.0,0.7467);
			rgb=(0.0,0.0,0.7867);
			rgb=(0.0,0.0,0.8267);
			rgb=(0.0,0.0,0.8667);
			rgb=(0.0,0.0933,0.8667);
			rgb=(0.0,0.1867,0.8667);
			rgb=(0.0,0.28,0.8667);
			rgb=(0.0,0.3734,0.8667);
			rgb=(0.0,0.4667,0.8667);
			rgb=(0.0,0.4934,0.8667);
			rgb=(0.0,0.52,0.8667);
			rgb=(0.0,0.5467,0.8667);
			rgb=(0.0,0.5733,0.8667);
			rgb=(0.0,0.6,0.8667);
			rgb=(0.0,0.6133,0.8267);
			rgb=(0.0,0.6267,0.7867);
			rgb=(0.0,0.64,0.7467);
			rgb=(0.0,0.6534,0.7067);
			rgb=(0.0,0.6667,0.6667);
			rgb=(0.0,0.6667,0.64);
			rgb=(0.0,0.6667,0.6133);
			rgb=(0.0,0.6667,0.5867);
			rgb=(0.0,0.6667,0.56);
			rgb=(0.0,0.6667,0.5333);
			rgb=(0.0,0.6534,0.4266);
			rgb=(0.0,0.64,0.32);
			rgb=(0.0,0.6267,0.2133);
			rgb=(0.0,0.6133,0.1067);
			rgb=(0.0,0.6,0.0);
			rgb=(0.0,0.6267,0.0);
			rgb=(0.0,0.6533,0.0);
			rgb=(0.0,0.68,0.0);
			rgb=(0.0,0.7066,0.0);
			rgb=(0.0,0.7333,0.0);
			rgb=(0.0,0.76,0.0);
			rgb=(0.0,0.7867,0.0);
			rgb=(0.0,0.8133,0.0);
			rgb=(0.0,0.84,0.0);
			rgb=(0.0,0.8667,0.0);
			rgb=(0.0,0.8934,0.0);
			rgb=(0.0,0.92,0.0);
			rgb=(0.0,0.9467,0.0);
			rgb=(0.0,0.9733,0.0);
			rgb=(0.0,1.0,0.0);
			rgb=(0.1467,1.0,0.0);
			rgb=(0.2933,1.0,0.0);
			rgb=(0.44,1.0,0.0);
			rgb=(0.5866,1.0,0.0);
			rgb=(0.7333,1.0,0.0);
			rgb=(0.7733,0.9867,0.0);
			rgb=(0.8133,0.9733,0.0);
			rgb=(0.8533,0.96,0.0);
			rgb=(0.8933,0.9466,0.0);
			rgb=(0.9333,0.9333,0.0);
			rgb=(0.9466,0.9066,0.0);
			rgb=(0.96,0.88,0.0);
			rgb=(0.9733,0.8533,0.0);
			rgb=(0.9867,0.8267,0.0);
			rgb=(1.0,0.8,0.0);
			rgb=(1.0,0.76,0.0);
			rgb=(1.0,0.72,0.0);
			rgb=(1.0,0.68,0.0);
			rgb=(1.0,0.64,0.0);
			rgb=(1.0,0.6,0.0);
			rgb=(1.0,0.48,0.0);
			rgb=(1.0,0.36,0.0);
			rgb=(1.0,0.24,0.0);
			rgb=(1.0,0.12,0.0);
			rgb=(1.0,0.0,0.0);
			rgb=(0.9733,0.0,0.0);
			rgb=(0.9467,0.0,0.0);
			rgb=(0.92,0.0,0.0);
			rgb=(0.8934,0.0,0.0);
			rgb=(0.8667,0.0,0.0);
			rgb=(0.8534,0.0,0.0);
			rgb=(0.84,0.0,0.0);
			rgb=(0.8267,0.0,0.0);
			rgb=(0.8133,0.0,0.0);
			rgb=(0.8,0.0,0.0);
			rgb=(0.8,0.16,0.16);
			rgb=(0.8,0.32,0.32);
			rgb=(0.8,0.48,0.48);
			rgb=(0.8,0.64,0.64);
			rgb=(0.8,0.8,0.8);
		}
}

 \newcommand{\KHTwoDPressureShortFig}[6]{

	\begin{tikzpicture}
		\renewcommand*\showkeyslabelformat[1]{}
		\makeatletter
		\def\SK@@ref#1>#2\SK@{}
		\makeatletter

		\begin{groupplot}[
				legend cell align={left},
				group style={
						group name=topLeft,
						group size=1 by 1,
						vertical sep=\plotSmallSeparation,
					},
				colormap name = viridis,
				view = {0}{90},
				xlabel={},
				xticklabel=\empty,
				ylabel={},
				yticklabel=\empty,
				width = \mylinewidth,
				height = \mylinewidth,
				colorbar,
				colorbar style={
						at={(-0.15,1)}, anchor=above north west, width=\plotBarWidth,
yticklabel pos=left,
						ylabel={Density $\rho_1$},
					},
				point meta min = 0.0,
				point meta max = 1.05,
				xmin=-0.5,
				xmax=0.5,
				ymin=-0.5,
				ymax=0.5,
			]

			\nextgroupplot[
				title={$t=1.3\times 10^{-2}$},
				title style={yshift=\plotGroupTitleRaise},
				legend to name={commonLegend},
				legend style={
						legend columns=2,
					},
			]

		\end{groupplot}

		\begin{groupplot}[
				legend cell align={left},
				group style={
						group name=topRight,
						group size=1 by 1,
						vertical sep=\plotSmallSeparation,
					},
				colormap name = viridis,
				view = {0}{90},
				xlabel={},
				xticklabel=\empty,
				ylabel={$y$},
				yticklabel pos=right,
				width = \mylinewidth,
				height = \mylinewidth,
point meta min = 0.0,
				point meta max = 1.05,
				xmin=-0.5,
				xmax=0.5,
				ymin=-0.5,
				ymax=0.5,
			]

			\nextgroupplot[
				title={$t=2.1\times 10^{-2}$},
				title style={yshift=\plotGroupTitleRaise},
				anchor=north west,
				at={($(topLeft c1r1.north east) + (\plotBigSeparation,0)$)},
			]

		\end{groupplot}

		\begin{groupplot}[
				legend cell align={left},
				group style={
						group name=midLeft,
						group size=1 by 1,
						vertical sep=\plotSmallSeparation,
					},
				colormap name = viridis,
				view = {0}{90},
				xlabel={},
				xticklabel=\empty,
				ylabel={},
				yticklabel=\empty,
				width = \mylinewidth,
				height = \mylinewidth,
				colorbar,
				colorbar style={
						at={(-0.15,1)}, anchor=above north west, width=\plotBarWidth,
yticklabel pos=left,
						ylabel={Density $\rho_2$},
					},
				point meta min = 0.0,
				point meta max = 2.15,
				xmin=-0.5,
				xmax=0.5,
				ymin=-0.5,
				ymax=0.5,
			]

			\nextgroupplot[
				anchor=north west,
				at={($(topLeft c1r1.south west) + (0,-\plotBigSeparation)$)},
				title={},
				title style={yshift=\plotGroupTitleRaise},
				legend to name={commonLegend},
				legend style={
						legend columns=2,
					},
			]

		\end{groupplot}

		\begin{groupplot}[
				legend cell align={left},
				group style={
						group name=midRight,
						group size=1 by 1,
						vertical sep=\plotSmallSeparation,
					},
				colormap name = viridis,
				view = {0}{90},
				xlabel={},
				xticklabel=\empty,
				ylabel={$y$},
				yticklabel pos=right,
				width = \mylinewidth,
				height = \mylinewidth,
point meta min = 0.0,
				point meta max = 2.15,
				xmin=-0.5,
				xmax=0.5,
				ymin=-0.5,
				ymax=0.5,
			]

			\nextgroupplot[
				title={},
				title style={yshift=\plotGroupTitleRaise},
				anchor=north west,
				at={($(midLeft c1r1.north east) + (\plotBigSeparation,0)$)},
			]

		\end{groupplot}

		\begin{groupplot}[
				legend cell align={left},
				group style={
						group name=botLeft,
						group size=1 by 1,
						vertical sep=\plotSmallSeparation,
					},
				colormap name = viridis,
				view = {0}{90},
				xlabel={$x$},
ylabel={},
				yticklabel=\empty,
				width = \mylinewidth,
				height = \mylinewidth,
				colorbar,
				colorbar style={
						at={(-0.15,1)}, anchor=above north west, width=\plotBarWidth,
yticklabel pos=left,
						ylabel={Total Pressure $p$},
					},
				point meta min = 0.5,
				point meta max = 1.15,
				xmin=-0.5,
				xmax=0.5,
				ymin=-0.5,
				ymax=0.5,
			]

			\nextgroupplot[
				anchor=north west,
				at={($(midLeft c1r1.south west) + (0,-\plotBigSeparation)$)},
				title={},
				title style={yshift=\plotGroupTitleRaise},
				legend to name={commonLegend},
				legend style={
						legend columns=2,
					},
			]

		\end{groupplot}

		\begin{groupplot}[
				legend cell align={left},
				group style={
						group name=botRight,
						group size=1 by 1,
						vertical sep=\plotSmallSeparation,
					},
				colormap name = viridis,
				view = {0}{90},
				xlabel={$x$},
ylabel={$y$},
				yticklabel pos=right,
				width = \mylinewidth,
				height = \mylinewidth,
				point meta min = 0.5,
				point meta max = 1.15,
				xmin=-0.5,
				xmax=0.5,
				ymin=-0.5,
				ymax=0.5,
			]

			\nextgroupplot[
				title={},
				title style={yshift=\plotGroupTitleRaise},
				anchor=north west,
				at={($(botLeft c1r1.north east) + (\plotBigSeparation,0)$)},
			]

			\addplot3 [
				surf,
				mesh/ordering=y varies,
				shader=flat,
			] table
				{\detokenize{#6}};

		\end{groupplot}

	\end{tikzpicture}

} %
  \newcommand\eps\varepsilon

\renewcommand{\O}{\mathcal{O}}

\newcommand{\curlyH}{\mathcal{H}}
\newcommand{\curlyS}{\mathcal{S}}

\newcommand{\curlyD}{\mathcal{D}}

\newcommand{\p}{_{p}}
\newcommand{\q}{_{q}}
\newcommand{\pq}{_{p,\,q}}

\usepgfplotslibrary{patchplots}

\pgfplotsset{every axis/.append style={
			grid=both,
			grid style={white, line width=.1pt},
			major grid style={white, line width=0.5pt},
			axis background/.style={fill=gray!10},
			axis line style={draw=none},
			tick style={draw=none},
			xlabel = $x$,
line width=1pt,
legend style={
					line width = 1pt,
					draw=none,
					/every even column/.append style={column sep=0.5cm}
				},
		}}

\usepgfplotslibrary{groupplots}

\usepackage{calc}

\definecolor{gg0}{HTML}{E24A33}
\definecolor{gg1}{HTML}{348ABD}
\definecolor{gg2}{HTML}{988ED5}
\definecolor{gg3}{HTML}{777777}
\definecolor{gg4}{HTML}{FBC15E}
\definecolor{gg5}{HTML}{8EBA42}
\definecolor{gg6}{HTML}{FFB5B8}

\pgfplotsset{
	/pgfplots/colormap={bright}{rgb255=(0,0,0) rgb255=(78,3,100) rgb255=(2,74,255)
			rgb255=(255,21,181) rgb255=(255,113,26) rgb255=(147,213,114) rgb255=(230,255,0)
			rgb255=(255,255,255)}
}

\usepackage{siunitx}
\usepackage{chemformula}

\usepackage[nolist]{acronym}
 
\newcommand\rd\dd
\newcommand\jac\JAC

  \allowdisplaybreaks

\renewcommand{\review}[2]{}
\renewcommand{\creview}[3]{}
\renewcommand{\ntcreview}[3]{}

\renewcommand{\tableofcontents}{}
\renewcommand{\listofreviews}{}

\expandafter\def\csname ver@etex.sty\endcsname{3000/12/31}

\usepackage{autonum}
\makeatletter
\patchcmd{\autonum@saveEnvironmentSubcommands}
{(0,0)\begin}
{(0,0)\hfuzz=\maxdimen\begin}
{}{}
\makeatother

\AtBeginDocument{

}

\makeatletter
\autonum@generatePatchedReferenceCSL{ref}
\autonum@generatePatchedReferenceCSL{eqref}
\autonum@generatePatchedReferenceCSL{Cref}
\makeatother

\newcommand*\showkeyslabelformat[1]{} 

\newcommand{\revision}[2]{#2}
\newcommand{\revisionNote}[2]{}

\definecolor{revisionColourOne}{RGB}{200,0,0}
\definecolor{revisionColourTwo}{RGB}{0,0,200}

\newcommand{\revisionOne}[1]{\revision{revisionColourOne}{#1}}
\newcommand{\revisionTwo}[1]{\revision{revisionColourTwo}{#1}}

\newcommand{\JAC}[1]{\creview{JAC}{red}{#1}}

\usepackage{siunitx}
\begin{document}

\begin{singlespace}\maketitle\end{singlespace}
\begin{abstract}
	We introduce an extension of the \acf{PIC} method that captures the Landau collisional effects in the \acl{VML} equations. The method arises from a regularisation of the variational formulation of the Landau equation, leading to a discretisation of the \revisionTwo{collision} operator that conserves mass, charge, momentum, and energy, while \revisionOne{increasing} the (regularised) entropy. The collisional effects appear as a fully deterministic effective force, thus the method does not require any transport-collision splitting. The scheme can be used in arbitrary dimension, and for a general interaction, including the Coulomb case. We validate the scheme on scenarios such as the Landau damping, the two-stream instability, and the Weibel instability, demonstrating its effectiveness in the numerical simulation of plasma.
\end{abstract}

\subjectclassification{\subjectPDF}
 \tableofcontents
\listofreviews

\begin{acronym}
	\acro{AP}[AP]{asymptotic-preserving}
	\acro{CPIC}[C-PIC]{collisional particle-in-cell}
	\acro{DSMC}[DSMC]{direct simulation Monte Carlo}
	\acro{FP}[FP]{Fokker-Planck}
	\acro{LFP}[LFP]{Landau-Fokker-Planck}
	\acro{MLMC}[MLMC]{multi-level Monte Carlo}
	\acro{PIC}[PIC]{particle-in-cell}
	\acro{RFP}[RFP]{Rosenbluth-Fokker-Planck}
	\acro{VA}[VA]{Vlasov-Amp\`{e}re}
	\acro{VAL}[VAL]{Vlasov-Amp\`{e}re-Landau}
	\acro{VL}[VL]{Vlasov-Landau}
	\acro{VM}[VM]{Vlasov-Maxwell}
	\acro{VML}[VML]{Vlasov-Maxwell-Landau}
	\acro{VP}[VP]{Vlasov-Poisson}
	\acro{VPL}[VPL]{Vlasov-Poisson-Landau}
	\acro{WLOG}[WLOG]{without loss of generality}
\end{acronym}
 
\section{Introduction}

This work introduces a fully deterministic extension of the \acf{PIC} method for the \acl{VM} equations that incorporates the effects of Landau collisions. The extension is based on a regularisation of the entropic structure of the Landau operator, and is able to preserve mass, charge, momentum, and energy, while mimicking the entropy \revisionOne{increase} structure of the problem. The method can be applied in arbitrary dimension and includes all particle interaction types, including the Coulomb interaction, relevant for plasmas and nuclear fusion.

The evolution of the electrons in a plasma can be modelled by the \acl{VL} equation:
\begin{align}\label{eq:vlasov}
	\pt f + v \cdot \gradx f + a(t,x,v) \cdot \gradv f = Q[f,f], \quad x\in \Omega \subset \R^\dimx, v\in \R^\dimv,
\end{align}
where $f=f(t,x,v)$ is the \textit{number distribution function} of the electrons in phase space. In the most complete models, the equation is posed in three physical dimensions ($\dimx=\dimv=3$), and the acceleration experienced by the electrons is derived from the \textit{Lorentz force}:
\begin{align}
	a(t,x,v) = \frac{q}{m} \prt*{E(t,x) + v\times B(t,x)},
\end{align}
where $m>0$ and $q<0$ are respectively the mass and charge of the electron.

The long-range interactions between charged particles are described by the \textit{electric} and \textit{magnetic fields}, $E$ and $B$, which satisfy \textit{Maxwell's equations}:
\begin{align} \label{eq:maxwell}
	\varepsilon_0 \mu_0 \pt E = \curlx B - \mu_0 J,\quad
	\pt B = -\curlx E,\quad
	\varepsilon_0 \divx E = \rho + \rho\ion,\quad
	\divx B = 0,
\end{align}
where $\varepsilon_0$ and $\mu_0$ are the \textit{permittivity} and \textit{permeability of free space}, and are related to the \textit{speed of light} by \revisionTwo{$c^2 = \prt*{\varepsilon_0 \mu_0}^{-1}$}. We assume the \textit{background ion density} $\rho\ion$ is a given positive constant. The \textit{charge density}, $\rho$, and the \textit{current density}, $J$, are defined as velocity moments of the distribution $f$,
\begin{align}
	\rho = q\int_{\R^\dimv} f \dv \quad\text{and}\quad J= q\int_{\R^\dimv} vf \dv.
\end{align}

The short-range interactions between electrons are described by $Q$, the \textit{Landau collision operator}:
\begin{align} \label{eq:landau}
	Q[f,f](v)=\divv \int_{\R^\dimv} A(v-\vst) \brk*{ f(\vst) \gradv f(v) - f(v) \gradvst f(\vst) } \dvst.
\end{align}
The \textit{collisional cross-section} $A$ is a symmetric and semi-positive-definite matrix given by
\begin{align}\label{eq:landau_matrix}
	\revisionOne{
		A(z) = \mathcal{C}_{\gamma} \abs{z}^{\gamma+2} \Pi(z),
		\quad \Pi(z) = \prt*{
			I_\dimv - \frac{z\otimes z}{\abs{z}^2}
		},
	}
\end{align}
\revisionOne{where $\mathcal{C}_{\gamma}>0$ is the \textit{collision strength}}. The matrix $\Pi(z)$ is the projection matrix onto the perpendicular of $z$, and $I$ is the $\dimv\times\dimv$ identity matrix. The exponent $\gamma$ determines the type of interaction, and is chosen in the range $-\dimv-1\leq\gamma\leq 1$ so that the expression in \eqref{eq:landau} is integrable. The most physically relevant choice for plasma is $\gamma=-\dimv=-3$, the \textit{Coulomb interaction}; \revisionOne{in this case, $\mathcal{C}_{-3} = \abs{\log\delta} 8^{-1} \pi^{-1} \varepsilon_0^{-2} m^{-2} q^4$, where $\log\delta$ is the so-called \textit{Coulomb logarithm}.}

The Landau collision operator is sometimes referred to as the \acl{LFP} operator, since it may be rewritten as a non-linear and non-local \acl{FP} operator:
\begin{align}\label{eq:landau operator}
	Q[f,f] = \divv \prt{A_f \gradv f - f a_f},
\end{align}
where the diffusion matrix $A_f$ and the drift $a_f$ are given by
\begin{align}
	A_f(t,v) = \int_{\R^\dimv} A(v-\vst) f(\vst) \dvst
	\quad\text{and}\quad
	a_f(t,v) = \int_{\R^\dimv} A(v-\vst) \gradvst f(\vst) \dvst.
\end{align}
In this form, sometimes also known as the \acl{RFP} operator, we see that the Landau operator is akin to a singular, non-local, and non-linear diffusion operator.

The numerical methods for collisionless plasma (the \acl{VM} equations) can be categorised as \acf{PIC} methods, finite difference/volume/element methods, and semi-Lagrangian methods; see \cite{Sonnendrucker2013} for a review. \ac{PIC} methods \cite{BL2018,HE1988} have often been favoured because they generally scale better, in view of the high dimensionality of the problem. Modern \ac{PIC} methods can be designed to preserve certain structural properties; for example, GEMPIC \cite{KKM2017} captures well the long-time behaviour of the equation by employing a symplectic integrator; on the other hand, the exact conservation of energy can be achieved using implicit time-stepping \cite{ML2011,CCB2011,Lapenta2017,RC2020,KS2021}, often exploiting a specific discretisation of Maxwell's equations \cite{Yee1966,HS1999}. \revisionTwo{PIC and grid-based methods exist with simultaneous energy and local charge conservation properties \cite{CCB2011,TC2015,TKC2015,CCY2020}.}

The main difficulty in the numerical simulation of collisionless plasma is the handling of the small-scale effects that arise from the electromagnetic fields, as they often lead to fine structure filamentation in phase space. Adding collisional effects will ease this difficulty while introducing a bigger one: the curse of dimensionality. Ultimately, the discretisation of the diffusive terms in six dimensions is very challenging from a computational efficiency perspective.

Several numerical methods have been proposed to discretise the Landau collision operator \eqref{eq:landau}, both in the spatially homogeneous and inhomogeneous settings. We classify them here as deterministic and stochastic methods. Among the deterministic class, in the homogeneous setting, we highlight the classical entropy schemes \cite{DL1994,BC1998,CF2004}, which preserve the conservation and dissipation properties of the Landau equation, and thus lead to the correct Maxwellian stationary states. Methods that discretise the Landau operator in the form \eqref{eq:landau operator} \revisionTwo{are given in \cite{TCS2015,TCS2016b,TCS2021}}. Implicit and \ac{AP} methods based on these entropy schemes have been developed in \cite{LM2005,JY2011}. We refer the reader to the review \cite{DP2014}.

Yet, the efficient approximation of the Landau operator remains a major challenge, even in the spatially homogeneous setting. The non-local nature of the collision operator leads to a quadratic complexity $\O(N^2)$, where $N$ is the number of discretisation elements. Faster algorithms that attempt to reduce this cost to $\O(N\log N)$ have been proposed, including multigrid algorithms \cite{BCD1997}, \revisionTwo{finite-volume methods \cite{TCS2016b, TCS2015}}, fast multipole expansions \cite{Lemou1998,Lemou2004}, and Fourier spectral methods \cite{PRT2000}; the latter deserves special attention, since it reduces the complexity to $\O(N\log N)$ through the fast Fourier transform by exploiting the convolutional properties of the Landau operator. The spectral method has been coupled with discretisation methods for the transport part to treat the inhomogeneous problems \cite{FP2002,DLP2015,ZG2017,HJS2018,LRW2021}; see \cite{Gamba2017} for a review. We also refer to semi-Lagrangian techniques \cite{KHC2016}.

\revisionTwo{Nevertheless, stochastic or Monte Carlo approaches remain the most frequently used tools among practitioners to simulate Coulomb collisions in an inhomogeneous setting, and we highlight two major classes.} The first is based on binary collisions \cite{TA1977,Nanbu1997}, in the spirit of the \ac{DSMC} method developed by Bird \cite{Bird1994} for the Boltzmann equation; see \cite{BN2000,CWD2008,DCP2010} for generalisations and \cite{MPZ2023} for a recent application in uncertainty quantification. The second is based on the drift-diffusion formulation of the Landau operator given in \eqref{eq:landau operator} \cite{MLJ1997,DCC2013}; see also \cite{RRD2014} for a multi-level Monte Carlo extension. The advantages and drawbacks of these stochastic approaches are well known: they are physically motivated and easy to implement, but their convergence is slow, they require many realisations due to statistical noise, and they cannot preserve the entropy structure of the problem.

The design of more efficient discretisation techniques for the Landau operator is a matter of utmost practical relevance, as many of the methods described above are prohibitively costly and cannot be used directly in real-world applications, such as the design and on-line control of modern nuclear fusion reactors. Mature plasma simulation tools such as NESO \cite{TAA2023} or XGC1 \cite{CK2008,KCD2009} typically include collisional effects through Monte Carlo approaches or mesh-based methods. The use of surrogate models trained on synthetic data for the collisional operator has recently been proposed \cite{MCD2021}.

This work introduces the \textit{\ac{CPIC} method}; a deterministic extension of \ac{PIC} methods also able to approximate the Landau collisional effects in the \acl{VML} equations. To construct the method, we generalise a recent particle approximation of the Landau operator introduced by \cite{CHW2020} in the homogeneous setting. The method exploits the variational properties of the Landau operator to propose a regularised entropic structure which is naturally discretised by particles, in a way that is fully compatible with any \ac{PIC} method. The \revisionOne{increment} of the (regularised) entropy is preserved at the discrete level, as are the conservations of mass, charge, momentum, and energy.

This approach has proven robust and flexible in the homogeneous setting, and has already been used for uncertainty quantification in the Landau equation \cite{BCM2023} and for the multispecies Landau equation \cite{ZPH2022,CHV2023}. The analysis of the homogeneous method and its convergence has been presented in \cite{CDW2022,CDW2023,CDD2024}. Moreover, a random batch technique was introduced in \cite{CJT2022} that can considerably reduce the computational cost of the method, while retaining all structural properties.

The generalisation to the inhomogeneous setting presented in this work presents a crucial departure from the previous works: the entropic structure is now regularised at a \textit{global} level, and the resulting regularised Landau operator delocalises in space. Since the particle approximation of the operator is performed in phase space, no splitting of transport and collisions is required; the collisional effects appear simply as an \textit{effective force}, alongside the Lorentz force of the \ac{PIC} approach. There is therefore no stochasticity and no splitting in our scheme. Moreover, the spatial structure, combined with the random batch approach, can be leveraged for an efficient implementation of the method that performs comparably to classical \ac{PIC} methods.

The rest of this work is organised as follows. In \cref{sec:method_sec}, we recall the physical properties and variational structure of the \acl{VML} equations, introduce the \ac{CPIC} method, and discuss its properties. In \cref{sec:experiments}, we perform a range of numerical simulations ($\dimx=1$ and $\dimv=2$) to validate the method and demonstrate its effectiveness, including explorations of the collisional effects on the Landau damping, the two-stream instability, and the Weibel instability. We conclude in \cref{sec:conclusion}, where we also present the outlook of this work.
 
\section{The Collisional Particle-In-Cell Method}\label{sec:method_sec}

This section introduces the \acf{CPIC} method and discusses its properties.

\subsection{The Vlasov-Maxwell-Landau Equations}\label{sec:VML}

The \acf{VML} equations \eqref{eq:vlasov}-\eqref{eq:maxwell}-\eqref{eq:landau} may, after non-dimensionalisation, be written as
\begin{subequations}\label{eq:continuity}
\begin{align+}
\label{eq:cont_vlasov} & \pt f + v \cdot \gradx f + \prt*{E+v\times B} \cdot \gradv f = Q[f,f],
\quad x \in \Omega\subseteq \R^\dimx, v \in \R^\dimv,
\\
\label{eq:cont_ampere_faraday}
& \revisionTwo{ \pt E = \nabla_x \times B - J, }
\quad \revisionTwo{ \pt B=-\nabla_x\times E, }
\\
\label{eq:cont_poisson_gauss}
& \revisionTwo{ \nabla_x \cdot E = \rho-\rho\ion, }
\quad \revisionTwo{ \nabla_x \cdot B=0, }
\\ \label{eq:cont_landau_1}
& Q\brk{f,f} = \divv\prt*{f\U\brk{f}},
\\ \label{eq:cont_landau_2}
& \U\brk{f}\prt*{x,v} = \int_{\R^\dimv} A \prt{v - \vst} b \brk{f} \prt{x,v,\vst} f \prt{x,\vst} \dvst,
\\ \label{eq:cont_landau_3}
& b \brk{f} \prt{x,v,\vst} = \gradv \vder{H}{f} \brk{f} \prt{x,v} - \gradvst \vder{H}{f} \brk{f} \prt{x,\vst},
\\ \label{eq:cont_landau_4}
& \revisionOne{H \brk{f}(x) = -S \brk{f}(x),}
\\ \label{eq:cont_landau_5}
& \revisionOne{S \brk{f}(x) = -\int_{\R^\dimv} f \log f \dv.}
\end{align+}
\revisionOne{Details of the non-dimensionalisation are given in the Appendix. Note that some terms have changed sign because the negative fundamental charge $q$ has been normalised.}
We will only consider quadrangular domains $\Omega$ equipped with periodic boundary conditions. \revisionOne{The cross-section matrix $A$ is the non-dimensionalised version of \eqref{eq:landau_matrix}}:
\begin{align+}\label{eq:landau_matrix_scaled}
A(z) = C \abs{z}^{\gamma+2} \Pi(z),
\quad \Pi(z) = \prt*{
	I_\dimv - \frac{z\otimes z}{\abs{z}^2}
},
\end{align+}
where $C>0$ is the \revisionOne{\textit{dimensionless collision strength}}, $\Pi(z)$ is the projection matrix onto $z^\perp$, and $I_\dimv$ is the identity matrix in $\dimv$ dimensions. By construction, the matrix $A$ is positive semi-definite.
\end{subequations}

\Cref{eq:continuity} possesses a wealth of physical properties: it conserves the mass, charge, momentum, and total energy of the system. It also \revisionOne{causes the \textit{thermodynamic entropy} \eqref{eq:cont_landau_5} to increase globally}, a property often called the \textit{H-Theorem}. \revisionOne{The name ``H-Theorem'' refers to \eqref{eq:cont_landau_4}, the \textit{information-theoretical entropy} $H$, which differs from $S$ by a sign. In the sequel, when we refer to \textit{entropy}, we refer to $S$.} We shall briefly recall these classical results for the sake of a complete exposition.

The Landau collision operator \eqref{eq:cont_landau_1} can be written in weak form as
\begin{align}\label{eq:weak_form}
	\int_{\R^\dimv} g(v) Q\brk{f,f} \dv
	 & = - \frac{1}{2} \iint_{\R^\dimv \times \R^\dimv} \prt*{ \gradv g(v) - \gradvst g(\vst) } \cdot A \prt{v - \vst} b \brk{f} \prt{v,\vst} f f_* \dv \dvst;
\end{align}
$f$ and $f_*$ stand respectively for $f(v)$ and $f(\vst)$, and the local dependence in $x$ has been omitted for simplicity. The expression vanishes for $g=1$, $g=v$, and $g=\frac{1}{2}\abs{v}^2$; these correspond, respectively, to the conservation of mass (and charge), momentum, and kinetic energy. The result is immediate for mass and momentum; for kinetic energy, one uses the fact that $A$ projects $b$ onto the perpendicular to $v-\vst$. Summarising,
\begin{align}\label{eq:operator_invariants}
	\int_{\R^\dimv} Q\brk{f,f} \dv = 0,\quad
	\int_{\R^\dimv} v Q\brk{f,f} \dv = 0,\quad
	\int_{\R^\dimv} \frac{v^2}{2} Q\brk{f,f} \dv = 0.
\end{align}
The weak form is also useful to prove the H-theorem: one chooses the test function $g=\vder{H}{f}=\log f+1$ in order to arrive at
\begin{align}\label{eq:operator_dissipation}
	\int_{\R^\dimv} \prt{\log f+1} Q\brk{f,f} \dv
	 & = - \frac{1}{2} \iint_{\R^\dimv \times \R^\dimv} b \brk{f} \prt{v,\vst} \cdot A \prt{v - \vst} b \brk{f} \prt{v,\vst} f f_* \dv \dvst \leq 0,
\end{align}
where the sign follows from the positive-semi-definite property of the matrix $A$.

A corollary to these properties is that the kernel of $Q$ is precisely the set of Maxwellian distributions parametrised by density, momentum, and energy, as is the case for the Boltzmann equation \cite{Villani1998,GZ2017}.

At the level of the full \ac{VML} system, the \revisionOne{increment} of the total entropy,
\begin{align}
	\revisionOne{
		\curlyS[f] \coloneqq \int_\Omega S[f](x) \dx = -\iint_{\Omega \times \R^\dimv} f \log f \dv \dx,
	}
\end{align}
is shown by multiplying the Vlasov equation \eqref{eq:cont_vlasov} by the test function $g=\vder{\curlyH}{f}=\log f+1$, integrating over the phase space, and using the boundary conditions/behaviour at infinity of $f$ to arrive at
\begin{align}\label{eq:continuous_dissipation}
	\revisionOne{
		\der{}{t} \curlyS[f] = -\iint_{\Omega\times\R^\dimv} \prt{\log f+1} Q\brk{f,f} \dv \dx \geq 0,
	}
\end{align}
using the H-theorem.

The global conservation properties are similarly derived by multiplying \eqref{eq:cont_vlasov} by the correct test function and integrating. Mass ($g=1$) immediately leads to
\begin{align}
	\der{}{t} \iint_{\Omega\times\R^\dimv} f \dv \dx = 0.
\end{align}
Momentum ($g=v$) results in
\begin{align}
	\der{}{t} \iint_{\Omega\times\R^\dimv} v f \dv \dx = \int_{\Omega} \prt*{\rho E + J\times B} \dx;
\end{align}
a classical computation involving Maxwell's equations leads to
\begin{align}
	\der{}{t} \prt*{
		\iint_{\Omega\times\R^\dimv} v f \dv \dx
		+ \int_\Omega E\times B \dx
	}
	= \int_{\Omega} \rho\ion E \dx;
\end{align}
in the case where the background ion density $\rho\ion$ is constant and there is no magnetic field, this quantity is exactly zero. Similarly, energy ($g=\frac{1}{2}\abs{v}^2$) leads to
\begin{align}
	\der{}{t} \frac{1}{2} \iint_{\Omega\times\R^\dimv} \abs{v}^2 f \dv \dx = \int_{\Omega} J\cdot E \dx;
\end{align}
once again, Maxwell's equations lead to
\begin{align}
	\der{}{t} \prt*{
		\frac{1}{2} \iint_{\Omega\times\R^\dimv} \abs{v}^2 f \dv \dx
		+ \frac{1}{2} \int_\Omega \prt*{\abs{E}^2 + \abs{B}^2} \dx
	}
	= 0
\end{align}
This computation is described in more detail in \cref{sec:global_properties}. \subsection{Description of the Method}\label{sec:method}

To devise the \acf{CPIC} method for \eqref{eq:continuity}, we shall seek an approximate particle solution of the form
\begin{align}\label{eq:particle_solution}
	\fN\prt{t,x,v}
	=
	\sum_{p=1}^{\Np}
	w\p
	\delta\prt*{x - x\p(t)}
	\delta\prt*{v - v\p(t)},
\end{align}
where $N$ is the number of particles, and where $w\p$, $x\p(t)$, and $v\p(t)$ are respectively the $p\th$ particle's weight, position, and velocity. Following the characteristics, we require $x\p(t)$ and $v\p(t)$ to solve
\begin{align}\label{eq:full_ode}
	\begin{cases}
		\displaystyle
		\der{x\p}{t} = v\p,
		\\[0.5em]
		\displaystyle
		\der{v\p}{t} = E\prt{t,x\p} + v\p \times B\prt{t,x\p} - \U\brk*{\fN}\prt*{x\p,v\p},
	\end{cases}
\end{align}
where $E\prt{t,x\p}$ and $B\prt{t,x\p}$ are the electromagnetic fields acting on the particle, and $\U\brk*{\fN}\prt*{x\p,v\p}$ is the \textit{effective force} arising from the collision term. In the absence of collisions, system \eqref{eq:full_ode} will reduce to the classical \ac{PIC} method for the \ac{VM} system \eqref{eq:vlasov}-\eqref{eq:maxwell}.

\subsubsection{Regularisations}

At the core of our scheme is an absolutely continuous regularisation of $\fN$. We will consider a \textit{spatial spline} $\Sx$, with scale parameters $\ex=\prt{\ex_1,\cdots,\ex_\dimx}$. The spline must satisfy the following properties:
\begin{enumerate}
	\item non-negativity: $\Sx(x)\geq 0$;
	\item symmetry: $\Sx(-x) = \Sx(x)$;
	\item unit mean: $\int_{\R^\dimx} \Sx(x) \dx = 1$.
\end{enumerate}
We will also consider a \textit{velocity spline} $\Sv$, with scale parameters $\ev=\prt{\ev_1,\cdots,\ev_\dimv}$, which must satisfy the same properties.

For the sake of concreteness, we choose the shape function
\begin{align}
	\revisionOne{G}(x)
	=
	\begin{cases}
		1 - \abs{x} & \text{if } \abs{x} \leq 1, \\
		0           & \text{otherwise},
	\end{cases}
\end{align}
and define the splines
\begin{align}
	 & \Sx(x) \coloneqq \frac{1}{\ex^\dimx} \Shx(x),
	\quad \Shx(x) \coloneqq \prod_{d=1}^{\dimx} \revisionOne{G}\prt*{\frac{x_d}{\ex_d}},
	\\&\Sv(x) \coloneqq \frac{1}{\ev^\dimv} \Shv(v),
	\quad \Shv(v) \coloneqq \prod_{d=1}^{\dimv} \revisionOne{G}\prt*{\frac{v_d}{\ev_d}}.
\end{align}
Here, $\ex^\dimx$ stands for the product $\prod_{d=1}^{\dimx}\ex_d$, and similarly $\ev^\dimv$ stands for $\prod_{d=1}^{\dimv}\ev_d$. We also use the convention $x=\prt{x_1,\cdots,x_\dimx}$ and $v=\prt{v_1,\cdots,v_\dimv}$. This choice of splines is practical, not only because of its simplicity, but also because the compact support of the splines can be exploited for computational efficiency (see \cref{sec:cellList}). However, we remark that the properties discussed in \cref{sec:properties,sec:global_properties} do not rely on this choice, only on the three properties discussed above.

Given a suitable choice of splines, we can define the \textit{regularised solution}
\begin{align}\label{eq:regularised_solution}
	\ftN\prt{t,x,v}
	\coloneqq
	\prt*{\Sx\Sv \convxv \fN}\prt{t,x,v}
	=
	\sum_{p=1}^{\Np}
	w\p
	\Sx\prt*{x - x\p(t)}
	\Sv\prt*{v - v\p(t)},
\end{align}
where ``$\convxv$'' denotes the double convolution in $x$ and $v$.

\subsubsection{Lorentz Term}\label{sec:lorentz_term}

In order to determine the electromagnetic force experienced by each particle, we must define the terms $E\prt{t,x\p}$ and $B\prt{t,x\p}$. To do so, information from the particles is first interpolated to a spatial mesh, where Maxwell's equations are solved. That solution is then extrapolated back to the particles.

\revisionTwo{
	The first step is to define a regularised electric charge current, specified as an integral of $\ftN$:
	\begin{align}
		\label{eq:current}
		\tilde J\prt{t,x}
		 & \coloneqq
		\int_{\R^\dimv} v \ftN\prt{t,x,v} \dv
		=
		\sum_{p=1}^{N} w\p v\p(t) \Sx\prt{x - x\p(t)}.
	\end{align}
	This can be evaluated on a spatial mesh $\Omega\h = \set{x\h}\h$ (assumed here to be a tensorised grid with mesh size $h_d$ on each dimension), yielding $\tilde J\prt{t,x\h}$.} These mesh values will be used as inputs to solve Maxwell's equations, yielding the electromagnetic fields $\tilde E\prt{t,x\h}$ and $\tilde B\prt{t,x\h}$ at the mesh points.

\revisionTwo{In practice, we only solve Amp\`{e}re's and Faraday's equations \eqref{eq:cont_ampere_faraday}; Poisson's and Gauss' equations \eqref{eq:cont_poisson_gauss} are only solved once, in order to determine $\tilde E\prt{0,x\h}$ and $\tilde B\prt{0,x\h}$ in a way that is consistent with the datum $\ftN\prt{0,x,v}$.} For further particulars, see \cref{sec:maxwell}.

Once the values of the fields are known on the mesh, they are extrapolated back to the particle positions:
\begin{align}\label{eq:fields}
	E\prt{t,x\p}
	\coloneqq
	\sum_h \tilde E\prt{t,x\h} \Shx\prt{x\p - x\h}
	= \sum_h \tilde E\prt{t,x\h} \Sx\prt{x\p - x\h} \ex^\dimx,
	\\\label{eq:fieldsB}
	B\prt{t,x\p}
	\coloneqq
	\sum_h \tilde B\prt{t,x\h} \Shx\prt{x\p - x\h}
	= \sum_h \tilde B\prt{t,x\h} \Sx\prt{x\p - x\h} \ex^\dimx.
\end{align}
This defines the Lorenz force acting on the $p\th$ particle, $E\prt{t,x\p} + v\p \times B\prt{t,x\p}$. If the terms \cref{eq:fields,eq:fieldsB} are seen as a spline reconstruction over a mesh,
\begin{align}
	E\prt{t,x\p}
	=
	\sum_h \tilde E\prt{t,x\h} \Sxh\prt{x\p - x\h},
	\quad
	B\prt{t,x\p}
	=
	\sum_h \tilde B\prt{t,x\h} \Sxh\prt{x\p - x\h},
\end{align}
it becomes evident that the choice of mesh size $h_d=\ex_d$ is, in fact, the only reasonable one, for any other would not scale appropriately.

\begin{remark}[Staggered grids]
	While the presentation in this section places $\tilde J$, $\tilde E$, and $\tilde B$ on the same mesh for simplicity, this is not required. In particular, staggered grids such as Yee's lattice \cite{Yee1966,HS1999} can be used instead. \revisionTwo{This is exactly what is used here, see \cref{sec:maxwell} for the full implementation details.}
\end{remark}

\subsubsection{Collision Term}\label{sec:collision_term}

We now construct a regularised collisional velocity field $\U\rxv\brk{f}\prt*{x,v}$. By analogy with \eqref{eq:continuity}, we define, for an arbitrary distribution $f$,
\begin{subequations}\label{eq:regularisation}
\begin{align+}
\U\rxv\brk{f}\prt*{x,v}
& =
\iint_{\Omega\times\R^\dimv}
\Sx \prt{x - \xst}
A \prt{v - \vst}
b \rxv \brk{f} \prt{x,\xst,v,\vst}
f \prt{\xst,\vst}
\dvst \dxst
, \\
b \rxv \brk{f} \prt{x,\xst,v,\vst}
& =
\gradv \vder{\H\rxv}{f} \brk{f} \prt{x,v}
-
\gradvst \vder{\H\rxv}{f} \brk{f} \prt{\xst,\vst}
, \\
\label{eq:regularisation_entropy}
\revisionOne{\H\rxv \brk{f}}
& =
\revisionOne{
	-\S\rxv \brk{f}
	,\quad
	\S\rxv \brk{f}
	=
	-\iint_{\Omega\times\R^\dimv}
	f \log\prt*{\Sx\Sv \convxv f}
	\dv\dx
	,}
\end{align+}
\end{subequations}
where $A$ is the collision kernel \eqref{eq:landau_matrix_scaled}.

The departure from \eqref{eq:continuity} is twofold: first, system \eqref{eq:regularisation} is well-defined for discrete measures; second, the entropy and collision operators have been altered to include spatial dependencies. Note that the discrete collision operator has been defined in terms of the total entropy \revisionOne{$\S$}, rather than the function of space $H(x)$ that appears in the continuous Landau operator. The physical interpretation of \eqref{eq:regularisation} is that first we \textit{delocalise} in space, to consider collisions in a neighbourhood of each point $x$, and then we \textit{localise} the interactions in order to associate each particle with a distinct position. This regularised collision somewhat resembles the \textit{Enskog collision operator}, which has been studied in the context of the Boltzmann equation \cite{Villani2006}.

To define the collisional component of the scheme, we will evaluate $\U\rxv$ for $\fN$ at the particle coordinates $\prt{x\p,v\p}$. First, we note that the regularised \revisionOne{functional} \eqref{eq:regularisation_entropy} can be rewritten as
\begin{align}
	\H\rxv \brk{f}
	 & =
	\iint_{\Omega\times\R^\dimv}
	f \log\prt{\tilde f}
	\dv\dx,
\end{align}
using $\tilde f$ as shorthand for $\prt*{\Sx\Sv \convxv f}$. It is straightforward to compute
\begin{align}
	\vder{\H\rxv}{f} \brk{f} \prt{x,v}
	=
	\log\prt{\tilde f}
	+
	\prt*{
		\frac{f}{\,\tilde f\,}
		\convxv
		\prt{\Sx\Sv}
	}
\end{align}
and
\begin{align}
	\gradv
	\vder{\H\rxv}{f} \brk{f} \prt{x,v}
	=
	\frac{\gradv \tilde f}{\tilde f}
	+
	\prt*{
		\frac{f}{\,\tilde f\,}
		\convxv
		\prt{\Sx\gradv\Sv}
	}.
\end{align}
We now structure the evaluation of the collision term in three sequential steps:
\begin{subequations}\label{eq:discrete_collision_operator}
\paragraph{Step I}
Compute the values of $\ftN$ and $\gradv\ftN$ at each particle:
\begin{align+}
\ftN\prt{x\p,v\p}
& =
\sum_{q=1}^{N}
w\q \Sx \prt{x\p - x\q} \Sv \prt{v\p - v\q}
, \\
\gradv \ftN\prt{x\p,v\p}
& =
\sum_{q=1}^{N}
w\q \Sx \prt{x\p - x\q} \gradv \Sv \prt{v\p - v\q}.
\end{align+}

\paragraph{Step II}
Compute the values of $\gradv \vder{\H\rxv}{f} \brk{\fN}$ at each particle:
\begin{align+}
\gradv \vder{\H\rxv}{f} \brk{\fN} \prt{x\p,v\p}
& =
\frac{\gradv \ftN\prt{x\p,v\p}}{\ftN\prt{x\p,v\p}}
+
\prt*{
	\frac{\fN}{\ftN}
	\convxv
	\prt{\Sx\gradv\Sv}
}
\prt{x\p,v\p}
\nonumber
\\
& =
\frac{\gradv \ftN\prt{x\p,v\p}}{\ftN\prt{x\p,v\p}}
+
\sum_{q=1}^{N}
w\q
\prt*{
	\frac{
		\Sx \prt{x\p - x\q} \gradv \Sv \prt{v\p - v\q}
	}{
		\ftN\prt{x\q,v\q}
	}
}
.
\end{align+}

\paragraph{Step III}
Compute the velocity field:
\begin{align+}\label{eq:step_3}
\U\rxv\brk{\fN}\prt{x\p,v\p}
& =
\sum_{q=1}^{N} w\q \Sx \prt{x\p - x\q} A\pq b\pq,
\end{align+}
where
\begin{align+}
A\pq
=
A \prt{v\p - v\q}
,
\quad
b\pq
=
\gradv \vder{\H\rxv}{f} \brk{\fN} \prt{x\p,v\p}
-
\gradv \vder{\H\rxv}{f} \brk{\fN} \prt{x\q,v\q}
.
\end{align+}
\end{subequations}
\Cref{sec:optimisation} discusses the practical implementation of these steps: the use of a cell list and a random batch approach can greatly reduce the computational cost of the method.

\begin{remark}
	An alternative regularisation of the \revisionOne{functional} is
	\begin{align}
		\iint_{\Omega\times\R^\dimv}
		\prt*{\Sx\Sv \convxv f} \log\prt*{\Sx\Sv \convxv f}
		\dv\dx,
	\end{align}
	reminiscent of the one employed in \cite{CHW2020} for the homogeneous Landau equation. Under this regularisation, the scheme will achieve very similar properties (see \cref{sec:properties,sec:global_properties}). However, the calculation of the corresponding $\gradv \vder{\H\rxv}{f} \brk{\fN} \prt{x\p,v\p}$ term will require the evaluation of an integral. The need for numerical quadrature makes this approach less convenient.
\end{remark}

\subsubsection{Choice of Regularisation Parameters}\label{sec:parameters}

The method involves two parameters, $\ex$ and $\ev$. However, it is not obvious how their value should be chosen, nor how it relates to the number of particles $N$. We present here a convention to make this choice more intuitive.

The computation of the Lorentz terms involves a spatial mesh, see \cref{sec:lorentz_term,sec:maxwell}. \Acl{WLOG}, we suppose the spatial domain is of the form $\Omega = \prt{0,L_1} \times \cdots \times \prt{0,L_\dimx}$ for $L_1,\cdots,L_\dimx > 0$. We assume that a uniform mesh is employed, and that each dimension is discretised with $N_{x_d}$ cells. We then choose
\begin{align}
	\ex_d \coloneqq \frac{L_d}{N_{x_d}} \text{ for each dimension }d.
\end{align}
The size of the spatial spline can therefore be understood as consequence of the \textit{spatial resolution} of the scheme.

\revisionTwo{We apply similar thinking for the velocity splines. The method does not require a mesh in velocity; however, we imagine a \textit{fictitious mesh}, and inspire a similar parameter choice. We stress, however, that this mesh is never used in computation, it is simply a thinking tool. Armed with this caveat, we approximate the velocity domain as $\prt{-L_v,L_v}^\dimv$, for some $L_v>0$. For each dimension, we choose a number of fictitious cells $N_{v_d}$, and let
	\begin{align}
		\ev_d \coloneqq \frac{L_v}{N_{v_d}} \text{ for each dimension }d.
	\end{align}
}

Given now the full mesh (space and velocity combined), we populate it with a number of \textit{particles per cell} $\Nc$. The total number of particles shall therefore be
\begin{align}
	N \coloneqq N_{x_1} \times \cdots \times N_{x_\dimx} \times N_{v_1} \times \cdots \times N_{v_\dimv} \times \Nc.
\end{align}
$\Nc$ is a measure of the \textit{granular resolution} of the scheme. Note that, while we speak of ``particles per cell'', the particles are not confined to individual cells and may roam the numerical domain. Therefore, the number of particles within a specific cell in phase space is not fixed \textit{a priori}.

To think of a notion of convergence of the scheme, we fix $\Nc$ (typically $\Nc=8$) and let $N_{x_d}$ and $N_{v_d}$ grow (see \cref{sec:validation}).
 \subsection{Computational Optimisation of the Method}\label{sec:optimisation}

In order to maximise the computational performance of the method, we describe two optimisations that we employ in our implementation of the scheme. The first, a cell list, does not alter the method. The second, the random batch method, does alter the scheme, but does not significantly affect its accuracy. In both cases, the structural properties of the scheme persist.

\subsubsection{Cell List Optimisation}\label{sec:cellList}

Cell lists \cite[Chapter~5.3.2]{AT1990} are data structures commonly used to efficiently find all pairs of particles within a given distance of each other. Their use in particle simulations can reduce the complexity of computing short-range particle interactions from $\O\prt{N^2}$ to $\O\prt{N \log N}$.

First, each particle is located within an auxiliary mesh (an $\O\prt{N}$ operation), and a list of which particle belongs in each mesh cell is stored. The size of the cells is chosen in relation to the size of the support of the splines, so that terms of the form $\Sx \prt{x\p - x\q}$ can only be non-zero if both particles lie in the same cell, or in cells which are neighbouring in space. Since the lists of particles in each cell are pre-computed, looking up each particle's ``neighbours'' is trivial. The same logic applies in the velocity dimensions, with the term $\Sv \prt{v\p - v\q}$.

The complexity of the three steps of \cref{sec:collision_term} reduces as follows:
\begin{enumerate}[I]
	\item This step involves $N$ particles with a compactly supported interaction; therefore, the evaluation cost is reduced from $\O\prt{N^2}$ to $\O\prt{N \log N}$.
	\item This step is as the previous one.
	\item This step involves an interaction which is local in space only (the Landau operator is, after all, non-local in velocity). Therefore, the cost reduction is lesser. The typical number of particles within one spatial cell is $\O\prt{N_{v_1} \times \cdots \times N_{v_\dimv} \times \Nc}$; the evaluation of the collision operator will be of quadratic order on this quantity. Accounting for the spatial cells, which do benefit from the localisation, the overall cost is estimated as
	      \begin{align}
		      \O\prt*{
			      N_{x_1} \times \cdots \times N_{x_\dimx} \times
			      \log\prt{N_{x_1} \times \cdots \times N_{x_\dimx}} \times
			      \prt{N_{v_1} \times \cdots \times N_{v_\dimv} \times \Nc}^2
		      },
	      \end{align}
	      which will be the dominant complexity of the numerical scheme.
\end{enumerate}

We remark that the cell list optimisation does not alter the method, as it only discards contributions which are exactly zero.

\subsubsection{Random Batch Optimisation}\label{sec:random_batch}

The random batch method \cite{JLL2020,CJT2022} is also a staple of particle simulations. Given the $N$ particles, we assign them randomly to $R$ \textit{batches} of equal size $N/R$. Steps I and II of \cref{sec:collision_term} are computed as before, but the collision operator of step III is replaced by
\begin{align}\label{eq:step_3_batch}
	\U\rxv\brk{\fN}\prt{x\p,v\p}
	 & =
	\frac{R\prt{N-1}}{N-R}
	\sum_{q\in {\mathcal{B}}_p} w\q \Sx \prt{x\p - x\q} A\pq b\pq,
\end{align}
where ${\mathcal{B}}_p$ is the batch which contains the $p\th$ particle. This term acts as an unbiased estimator of the original velocity field.

The complexity of the third step of \cref{sec:collision_term} is reduced by the batching. Since the batch size is $N/R$, the interaction is quadratic, and it must be computed for each batch, the overall cost is $\O\prt{N^2\times R^{-1}}$. If we combine this technique with the cell list, the resultant complexity is estimated as
\begin{align}
	\O\prt*{
		N_{x_1} \times \cdots \times N_{x_\dimx} \times
		\log\prt{N_{x_1} \times \cdots \times N_{x_\dimx}} \times
		\prt{N_{v_1} \times \cdots \times N_{v_\dimv} \times \Nc}^2 \times R^{-1}
	}.
\end{align}
This remains the dominant complexity of the numerical scheme.

The random batch optimisation does alter the numerical method. However, it is easy to prove that the structural properties of the method presented in \cref{sec:properties,sec:global_properties} survive the batching; see \cite{CJT2022} for details, where the random batch method was used to accelerate the particle method for the homogeneous Landau equation \cite{CHW2020}. Furthermore, the use of batches does not significantly affect the accuracy of the method, see \cref{sec:validation}.
 \subsection{Properties of the Regularised Collision Operator}\label{sec:properties}

This section discusses the structural properties of the discrete collision operator \eqref{eq:discrete_collision_operator}.

\subsubsection{Collision Invariants}

The conservation properties of the method arise naturally from the definition of the discrete collision operator \eqref{eq:discrete_collision_operator}. Specifically, we are able to emulate the weak form \eqref{eq:weak_form} of the Landau operator at the discrete level, which shows that the scheme preserves the usual collision invariants: $1$, $v$, and $\frac{1}{2}\abs{v}^2$.

The discrete weak form is easily found. By analogy with
\begin{align}
	\int_{\R^\dimv} g(v) Q\brk{f,f} \dv
	= \int_{\R^\dimv} g(v) \divv\prt*{f\U\brk{f}} \dv
	= -\int_{\R^\dimv} \gradv g(v) \cdot \prt*{f\U\brk{f}} \dv,
\end{align}
we compute, for a test function $g(x,v)$,
\begin{align}
	 & -\sum_{p=1}^{N} w\p \gradv g(x\p,v\p) \cdot \U\rxv\brk{\fN}\prt{x\p,v\p}
	\\&=
	-\sum_{p=1}^{N} w\p \gradv g(x\p,v\p) \cdot \sum_{q=1}^{N} w\q \Sx \prt{x\p - x\q} A\pq b\pq
	\\&=
	-\sum_{p,q=1}^{N} w\p w\q \Sx \prt{x\p - x\q} \gradv g(x\p,v\p) \cdot A\pq b\pq
	\\&=
	-\frac{1}{2}\sum_{p,q=1}^{N} w\p w\q \Sx \prt{x\p - x\q} \brk*{ \gradv g(x\p,v\p) - \gradv g(x\q,v\q)} \cdot A\pq b\pq,
\end{align}
symmetrising on the last line, and exploiting the symmetry $\Sx \prt{x\p - x\q}$ and the antisymmetry of $b\pq$. This term vanishes trivially for $g=1$ as well as $g=v$. To see that it also vanishes for $g=\frac{1}{2}\abs{v}^2$, one uses the fact that $A\pq$ projects $b\pq$ onto the perpendicular to $v\p - v\q$. We therefore conclude:

\begin{theorem}[Discrete collision invariants]\label{th:invariants}
	The functions $g=1$, $g=v$, and $g=\frac{1}{2}\abs{v}^2$ are collision invariants of the \ac{CPIC} collision operator:
	\begin{align}
		-\sum_{p=1}^{N} w\p \gradv g(x\p,v\p) \cdot \U\rxv\brk{\fN}\prt{x\p,v\p} = 0.
	\end{align}
\end{theorem}

\subsubsection{H-Theorem}

The discrete H-theorem in phase space also holds, and is readily obtained from the discrete weak form. At the continuous level, one would choose the test function $g=\log f+1=\vder{\H}{f}$. Instead, we choose $g=\vder{\H\rxv}{f} \brk{\fN}$, to find
\begin{align}
	\curlyD\rxv
	 & \coloneqq
	\sum_{p=1}^{\Np} w\p \gradv \vder{\H\rxv}{f} \brk{\fN} \prt{x\p,v\p} \cdot \U\rxv\brk{\fN}\prt*{x\p,v\p}
	\\&=
	\frac{1}{2}\sum_{p,q=1}^{\Np} w\p w\q \Sx \prt{x\p - x\q} \brk*{ \gradv \vder{\H\rxv}{f} \brk{\fN} \prt{x\p,v\p} - \gradv \vder{\H\rxv}{f} \brk{\fN} \prt{x\q,v\q}} \cdot A\pq b\pq
	\\&=
	\frac{1}{2}\sum_{p,q=1}^{\Np} w\p w\q \Sx \prt{x\p - x\q} b\pq \cdot A\pq b\pq,
\end{align}
which is non-positive due to the positive-semi-definite property of the matrix $A\pq$. We therefore conclude:

\begin{theorem}[Discrete H-theorem]\label{th:dissipation}
	The H-theorem holds for the \ac{CPIC} entropy and variation:
	\begin{align}
		\curlyD\rxv
		=
		\sum_{p=1}^{\Np} w\p \gradv \vder{\H\rxv}{f} \brk{\fN} \prt{x\p,v\p} \cdot \U\rxv\brk{\fN}\prt*{x\p,v\p}
		\leq 0.
	\end{align}
\end{theorem}

\subsection{\revisionTwo{Global} Properties of the \acs{CPIC} Method}\label{sec:global_properties}

This section discusses the properties of the full \ac{CPIC} method \eqref{eq:full_ode}. By construction, the discretisation of the collision operator preserves the conservation properties and H-theorem, as discussed in \cref{sec:properties}. Therefore, \ac{CPIC} inherits the properties of the underlying \ac{PIC} implementation.

\subsubsection{Conservation of Charge \& Energy}\label{sec:global_properties_conservation}

We may compute the global evolution in time of a test function $g(x,v)$:
\begin{align}
	 &
	\der{}{t} \iint_{\Omega\times\R^\dimv} g(x,v) \fN\prt{t,x,v} \dv\dx
= \der{}{t} \sum_{p=1}^{N} w\p g(x\p,v\p)
	\\&
	= \sum_{p=1}^{N} w\p \brk*{ \gradx g(x\p,v\p) \cdot \dot{x}\p + \gradv g(x\p,v\p) \cdot \dot{v}\p }
	\\&
	= \sum_{p=1}^{N} w\p \brk*{ \gradx g(x\p,v\p) \cdot v\p + \gradv g(x\p,v\p) \cdot \prt*{
			E\prt{t,x\p} + v\p \times B\prt{t,x\p} - \U\rxv\brk{\fN}\prt*{x\p,v\p}
		}}.
\end{align}
Mass and charge ($g=1$) are trivially conserved \revisionTwo{at the global level.}

The evolution of the kinetic energy ($g=\frac{1}{2}\abs{v}^2$) is given by
\begin{align}
	 &
	\der{}{t} \frac{1}{2} \iint_{\Omega\times\R^\dimv}\abs{v}^2 \fN\prt{t,x,v} \dv\dx
	=
	\der{}{t} \frac{1}{2} \sum_{p=1}^{N} w\p \abs{v\p}^2
	\\&
	= \sum_{p=1}^{N} w\p v\p \cdot \prt*{ E\prt{t,x\p} + v\p \times B\prt{t,x\p} - \U\rxv\brk{\fN}\prt*{x\p,v\p} }
	\\&
	= \sum_{p=1}^{N} w\p v\p \cdot E\prt{t,x\p}
	\\&
	= \sum_h \sum_{p=1}^{N} w\p v\p \cdot \tilde E\prt{t,x\h} \Sx\prt{x\p - x\h} h^\dimx
	\\&
	= \sum_h \tilde J \prt{t,x\h} \cdot \tilde E\prt{t,x\h} h^\dimx,
\end{align}
where we have used \cref{th:invariants}, the conservation of momentum at the level of the collision operator.
At the continuous level, the conservation of energy identity would be found by invoking Maxwell's equations in order to relate $J\cdot E$ to the electromagnetic energy. Namely:
\begin{align}
	\pt \prt*{ \frac{1}{2}\abs{E}^2 + \frac{1}{2}\abs{B}^2 }
	= E \cdot \pt E + B \cdot \pt B
	= -J \cdot E - \divx \prt{E \times B},
\end{align}
a computation that requires the vector identity
\begin{align}\label{eq:vector_identity}
	\divx (E\times B)=(\nabla_x\times E)\cdot B-E\cdot (\nabla_x \times B).
\end{align}

\begin{remark}[Conservation of energy in the fully discrete setting]
	\revisionTwo{
		Both the chain rule in time and the vector identity in space will not hold for a general discretisation; however, the correct approach is well-known. In space, Yee's lattice \cite{Yee1966,HS1999} leads to the corresponding discrete vector identities. In time, implicit schemes (the midpoint rule, in particular) can be used to preserve the chain rule \cite{ML2011,CCB2011,CEF2015,Lapenta2017,KS2021}. Under such conditions, we would arrive at the balance of energy:
		\begin{align}
			\der{}{t} \prt*{
				\mathcal{E}_K
				+\mathcal{E}_E
				+\mathcal{E}_B
			}
			\coloneqq
			\der{}{t} \prt*{ \frac{1}{2} \sum_{p=1}^{N} w\p \abs{v\p}^2 + \frac{1}{2}\sum_h \abs{\tilde E\prt{t,x\h}}^2 h^\dimx + \frac{1}{2}\sum_h \abs{\tilde B\prt{t,x\h}}^2 h^\dimx}
			\equiv 0.
		\end{align}
	}

	\revisionTwo{
		The numerical examples of \cref{sec:experiments} are discretised using Yee's lattice (see \cref{sec:maxwell}), but employ an explicit time integrator for the sake of efficiency. Therefore, energy is only conserved at the semi-discrete level (continuous in time), but generally not conserved at the fully discrete level.
	}
\end{remark}

\begin{remark}[Conservation of Momentum]
	A similar analysis can be performed for the conservation of momentum. Similarly, a specific chain rule in time and vector identity in space are required to hold in order to derive the correct conservation identity. In general, these will not hold, even in the discretisations that do conserve the energy. We are not aware of any discretisation capable of exactly conserving energy and momentum simultaneously.
\end{remark}

\revisionTwo{
	\begin{remark}[Local Conservation of Charge]
		The local conservation of charge equation,
		\begin{align}\label{eq:local_conservation_charge}
			\pt\rho + \div J = 0,
		\end{align}
		a continuity equation for the electric charge, can be derived from Maxwell's equations and also from Vlasov's equation. This equation is intimately related to the conservation of energy; as such, it is desirable that a PIC scheme should verify a discrete version of \eqref{eq:local_conservation_charge}. PIC methods which exhibit local charge conservation have been proposed in several works, including \cite{VB1992,Esirkepov2001,CCB2011,CCY2020}.
	\end{remark}
}

\subsubsection{\revisionOne{Increment} of Entropy}\label{sec:dissipation_entropy}

At the continuous level, the global \revisionOne{increment} of the entropy \eqref{eq:continuous_dissipation} is a consequence of two facts: the H-theorem, and the fact that the Vlasov equation conserves entropy. Since the discrete H-theorem has already been shown in \cref{sec:properties}, we must study the conservation.

The evolution of the regularised entropy is not \textit{a priori} given by the expression in \cref{sec:global_properties_conservation} since the test function, $g=\vder{\H\rxv}{f} \brk{\fN}$, also depends in time. However, a direct computation shows
\begin{align}
	 & \der{}{t} \revisionOne{\S\rxv} \brk{\fN}
	=
	\revisionOne{-} \der{}{t} \iint_{\Omega\times\R^\dimv} \fN \log\prt{\ftN} \dv\dx
	=
	\revisionOne{-} \der{}{t} \sum_{p=1}^{\Np} w\p \log\prt{\ftN(x\p,v\p)}
	\\&=
\revisionOne{-} \sum_{p,q=1}^{\Np} w\p w\q \frac{ \der{}{t} \brk*{ \vphantom{x^2} \Sx\prt*{x\p - x\q} \Sv\prt*{v\p - v\q} } }{ \ftN(x\p,v\p) }
	\\&=
	\revisionOne{-} \sum_{p,q=1}^{\Np} w\p w\q \frac{
		\gradx \Sx\prt*{x\p - x\q} \Sv\prt*{v\p - v\q} \cdot \prt*{\dot x\p - \dot x\q}
		+ \Sx\prt*{x\p - x\q} \gradv \Sv\prt*{v\p - v\q} \cdot \prt*{\dot v\p - \dot v\q}
	}{ \ftN(x\p,v\p) }
	.
\end{align}
The first summand can be greatly simplified, noting
\begin{align}
	 &
	\sum_{p,q=1}^{\Np} w\p w\q \frac{
		\gradx \Sx\prt*{x\p - x\q} \Sv\prt*{v\p - v\q} \cdot \prt*{\dot x\p - \dot x\q}
	}{ \ftN(x\p,v\p) }
\\&=
	\sum_{p,q=1}^{\Np} w\p w\q \frac{ \gradx \Sx\prt*{x\p - x\q} \Sv\prt*{v\p - v\q} }{ \ftN(x\p,v\p) } \cdot \dot x\p
	+
	\sum_{p,q=1}^{\Np} w\p w\q \frac{ \gradx \Sx\prt*{x\p - x\q} \Sv\prt*{v\p - v\q} }{ \ftN(x\q,v\q) } \cdot \dot x\p
	\\&=
	\sum_{p=1}^{\Np} w\p \brk*{
		\frac{ \gradx \ftN(x\p,v\p) }{ \ftN(x\p,v\p) }
		+
		\sum_{q=1}^{\Np} w\q \frac{ \gradx \Sx\prt*{x\p - x\q} \Sv\prt*{v\p - v\q} }{ \ftN(x\q,v\q) }
	} \cdot \dot x\p
	\\&=
	\sum_{p=1}^{\Np} w\p \brk*{ \frac{ \gradx \ftN(x\p,v\p) }{ \ftN(x\p,v\p) } + \prt*{ \frac{\fN}{\ftN} \convxv \prt{\Sx\gradv\Sv}} \prt{x\p,v\p} } \cdot \dot x\p
	\\&=
	\sum_{p=1}^{\Np} w\p \gradx \vder{\H\rxv}{f} \brk{\fN} \prt{x\p,v\p} \cdot \dot x\p
	=
	\sum_{p=1}^{\Np} w\p \gradx \vder{\H\rxv}{f} \brk{\fN} \prt{x\p,v\p} \cdot v\p,
\end{align}
where we have swapped the labels $p$ and $q$ in the second term of the second line, and have used the antisymmetry of $\gradx \Sx\prt*{x\p - x\q}$. The last expression is the discrete analogue of the term $\iint_{\Omega\times\R^\dimv} \prt*{\log f +1 } \divx \prt*{fv} \dv\dx$, which vanishes under reasonable assumptions. However, the discrete term is not zero in general; i.e. the \ac{PIC} method does not conserve the regularised entropy exactly. We define the \textit{entropy conservation error in position},
\begin{align}
	\mathcal{C}_x \coloneqq \revisionOne{-} \sum_{p=1}^{\Np} w\p \gradx \vder{\H\rxv}{f} \brk{\fN} \prt{x\p,v\p} \cdot v\p,
\end{align}
which will be shown to be numerically negligible.

The second summand in the evolution of $\H\rxv \brk{\fN}$ similarly becomes
\begin{align}
	 &
	\sum_{p=1}^{\Np} w\p \gradv \vder{\H\rxv}{f} \brk{\fN} \prt{x\p,v\p} \cdot \dot v\p
	\\&=
	\sum_{p=1}^{N} w\p \gradv \vder{\H\rxv}{f} \brk{\fN} \cdot \prt*{ E\prt{t,x\p} + v\p \times B\prt{t,x\p } - \U\rxv\brk{\fN}\prt*{x\p,v\p}
	}
	\\&=
	\sum_{p=1}^{N} w\p \gradv \vder{\H\rxv}{f} \brk{\fN} \cdot \prt*{ E\prt{t,x\p} + v\p \times B\prt{t,x\p} }
	- \curlyD\rxv
	\\&\leq
	\sum_{p=1}^{N} w\p \gradv \vder{\H\rxv}{f} \brk{\fN} \cdot \prt*{ E\prt{t,x\p} + v\p \times B\prt{t,x\p} },
\end{align}
using the discrete H-theorem (\cref{th:dissipation}). The remaining quantity is the discrete analogue of the continuous term $\iint_{\Omega\times\R^\dimv} \prt*{\log f +1 } \divv \prt*{f\prt*{E+v\times B}} \dv\dx$; once again, the continuous integral vanishes, but the discrete term does not, in general. We define the \textit{entropy conservation error in velocity},
\begin{align}
	\mathcal{C}_v \coloneqq \revisionOne{-}\sum_{p=1}^{N} w\p \gradv \vder{\H\rxv}{f} \brk{\fN} \cdot \prt*{ E\prt{t,x\p} + v\p \times B\prt{t,x\p} },
\end{align}
which will also be shown to be numerically negligible.

To summarise, we find
\begin{align}
	\der{}{t} \revisionOne{\S\rxv} \brk{\fN}
	= \mathcal{C}_x + \mathcal{C}_v \revisionOne{+\curlyD\rxv},
\end{align}
where $\revisionOne{\curlyD\rxv\geq 0}$. Because the \ac{PIC} method does not exactly conserve entropy, the \ac{CPIC} method cannot be said to exactly \revisionOne{increase} it. However, both $\mathcal{C}_x$ and $\mathcal{C}_v$ will be shown to be numerically negligible in \cref{sec:LandauDamping}.
 \subsection{Full Discretisation and Time Stepping}\label{sec:maxwell}

This section describes the full discretisation of the field solver and our time stepping scheme. We limit the implementation to the 1D-2V setting (one dimension in space, two dimensions in velocity; $\dimx=1$, $\dimv=2$) for the sake of computation, although the method as described in \cref{sec:method} also applies in higher dimensions.

We solve the \ac{VML} equations over a time interval $(0,T)$, where $T>0$, and discretise it with a uniform step $\Dt$. We describe here the update from time $t\n$ to $t\np$, where $t\n\coloneqq n\Dt$.

\subsubsection{Field Update}

In 1D-2V, Maxwell's equations reduce to
\begin{align}\label{eq:maxwell_1D2V}
	\pt E_1 = - J_1,
	\quad
	\pt E_2 = - J_2 - \px B_3,
	\quad
	\pt B_3 = - \px E_2,
	\quad x \in \Omega = (0,L),
\end{align}
for some $L>0$, using the convention $E=(E_1,E_2,0)$, $B=(0,0,B_3)$, $J=(J_1,J_2,0)$. We prescribe periodic boundary conditions in space.

In order to discretise \eqref{eq:maxwell_1D2V}, we choose a number of cells $N_x$, and compute the corresponding regularisation parameter $\eta=L/N_x$. We construct two meshes with spacing $\eta$; the spacing of the mesh must be related to the spatial regularisation parameter, see \cref{sec:lorentz_term,sec:parameters}. We define a \textit{primal mesh} $\Omega=\set*{x\i \mid 1\leq i\leq N_x}$ and a dual mesh $\Omega'=\set*{x\ih \mid 0\leq i\leq N_x}$, where $x\i \coloneqq (i-1/2) \eta$.

The field update begins with the particles and the electromagnetic field at time $t\n$. The positions and velocities of the particles are $x\p\n = (x_{1,\,p}\n)$ and $v\p\n = (v_{1,\,p}\n, v_{2,\,p}\n)$ for $1\leq p\leq N$. The electric field is considered on the primal mesh; the first component is $\tilde{E}_{1,\,i}\n$, and the second component is $\tilde{E}_{2,\,i}\n$, for $1\leq i\leq N_x$. The magnetic field is considered on the dual mesh; the third component is $\tilde{B}_{3,\,i}\n$ for $0\leq i\leq N_x$.

The current is evaluated over the primal mesh $\Omega$ through the interpolation described in \cref{sec:lorentz_term}, from $x\p\n$ and $v\p\n$, yielding $\tilde{J}_{1,\,i}\n$ and $\tilde{J}_{2,\,i}\n$ for $1\leq i\leq N_x$. Then, the field update is
\begin{align}
	 &
	\frac{ \tilde{E}_{1,\,i}\np - \tilde{E}_{1,\,i}\n }{\Dt}
	=
	-\tilde{J}_{1,\,i}\n,
	\quad\text{ for } 1\leq i\leq N_x,
	\\&
	\frac{ \tilde{E}_{2,\,i}\np - \tilde{E}_{2,\,i}\n }{\Dt}
	=
	- \tilde{J}_{2,\,i}\n
	- \frac{ \tilde{B}_{3,\,i+\nhalf}\n - \tilde{B}_{3,\,i-\nhalf}\n }{\Dx},
	\quad\text{ for } 1\leq i\leq N_x,
	\\&
	\frac{ \tilde{B}_{3,\,i+\nhalf}\np - \tilde{B}_{3,\,i+\nhalf}\n }{\Dt}
	=
	- \frac{ \tilde{E}_{2,\,i+1}\n - \tilde{E}_{2,\,i}\n }{\Dx},
	\quad\text{ for } 1\leq i\leq N_x;
\end{align}
the periodic boundary conditions are
\begin{align}
	\tilde{B}_{3,\,\nhalf}\n = \tilde{B}_{3,\,N_x+\nhalf}\n
	\quad\text{and}\quad
	\tilde{E}_{2,\,N_x+1}\n = \tilde{E}_{2,\,1}\n.
\end{align}
This spatial discretisation is Yee's lattice \cite{Yee1966,HS1999} applied to the 1D-2V setting, \revisionTwo{a choice motivated by conservation of energy considerations, see \cref{sec:global_properties_conservation}.}

To conclude, the fields acting on the particles, $E_{1,\,p}\np$, $E_{2,\,p}\np$, and $B_{3,\,p}\np$, are computed from $ \tilde{E}_{1,\,i}\np$, $\tilde{E}_{2,\,i}\np$, and $\tilde{B}_{3,\,i+\nhalf}\np$ using the extrapolation described in \cref{sec:lorentz_term}. In the numerical experiments without magnetic field, we simply set $\tilde{B}_{3,\,i+\nhalf}\equiv 0$.

\subsubsection{Particle Update}

Given the fields acting on the particles at time $t\n$, $E_{1,\,p}\n$, $E_{2,\,p}\n$, and $B_{3,\,p}\n$, as well as the collision term $\U\rxv\brk{\fN} = \prt*{ \U_{1,\,\ex,\,\ev}\brk{\fN}, \U_{2,\,\ex,\,\ev}\brk{\fN} }$ described in \cref{sec:collision_term}, the particle update is
\begin{align}
	 &
	\frac{ x_{1,\,p}\np - x_{1,\,p}\n }{\Dt} = v_{1,\,p}\np,
	\\&
	\frac{ v_{1,\,p}\np - v_{1,\,p}\n }{\Dt} = E_{1,\,p}\n + v_{2,\,p}\n B_{3,\,p}\n - 	\U_{1,\,\ex,\,\ev}\brk{\fN}\prt{x\p,v\p},
	\\&
	\frac{ v_{2,\,p}\np - v_{2,\,p}\n }{\Dt} = E_{2,\,p}\n - v_{1,\,p}\n B_{3,\,p}\n - 	\U_{2,\,\ex,\,\ev}\brk{\fN}\prt{x\p,v\p}.
\end{align}
 \subsection{Initialisation from Continuous Data}

This section discusses the initialisation of the \ac{CPIC} method \eqref{eq:full_ode}. We again limit the discussion to the 1D-2V setting, though our approach can be generalised to higher dimensions.

\subsubsection{Particle Initialisation}\label{sec:sampling}
Given an initial datum $f_0(x,v_1,v_2)$, the clear approach to particle initialisation is to construct the initial particle positions and velocities $x\p^0 = (x_{1,\,p}^0)$ and $v\p^0 = (v_{1,\,p}^0, v_{2,\,p}^0)$ as samples $\prt{x_{1,\,p}^0,v_{1,\,p}^0, v_{2,\,p}^0}$ from the distribution $f_0$.

Whenever possible, exact sampling tools are used. For instance, the marginals of $f_0$ in velocity are Gaussian mixtures in most of the numerical experiments entertained in \cref{sec:experiments}; in those cases, sampling the velocity coordinates $(v_{1,\,p}^0, v_{2,\,p}^0)$ is trivial.

In the cases where an exact sampling is not known (in \cref{sec:BKW}, and the marginal of $f_0$ in space of the rest of experiments), we use \textit{stratified sampling}. The support of the target distribution is discretised in a mesh (in our case, the spatial mesh described in \cref{sec:maxwell}), and particles are sampled uniformly on each mesh cell. The number of particles sampled on each cell is proportional to the integral of the target distribution on that cell (which we approximate numerically).

\paragraph{Symmetries}
It is possible to enforce certain symmetries in the particle sampling. In the test cases of \cref{sec:collisions}, we ensure that the initial momentum of the particle solution is zero. This is done by performing the sampling with $N/2$ particles, and then generating $N/2$ additional particles through a $\pi$ radians rotation about the velocity origin.

\subsubsection{Field Initialisation}

The numerical experiments of \cref{sec:experiments} all use self-consistent initialisations for the fields. Whenever present, the fields are initialised with approximations of $\divx E_0=\rho-\rho\ion$ and $\divx B_0=0$.
  
\section{Numerical Experiments}\label{sec:experiments}

In this section we validate the \ac{CPIC} method and demonstrate its effectiveness in dealing with a range of collisional plasma simulations. Interactive versions of the simulations presented in this section are available online \cite{BCH2024Web}. Videos of the simulations can be found in the permanent repository \cite{BCH2024Fig}.

\subsection{Validation}\label{sec:validation}

\subsubsection{Order of Convergence (2V)}\label{sec:BKW}

We begin the validation of the scheme by considering a spatially homogeneous problem. In this simplified setting, our scheme would be analogous to that of \cite{CHW2020}; however, our ``asymmetric'' regularisation of the entropy is precisely the choice not pursued in the reference. The validation is therefore novel. Furthermore, we are validating the random initialisation described in \cref{sec:sampling}, as well as the random batch approach described in \cref{sec:random_batch}.

We shall reproduce a test from \cite{CHW2020}, \revisionOne{which studies the relaxation to the Maxwellian in a homogeneous setting. The initial condition is a ring-like density,
	\begin{align}
		f_0(v_1,v_2)
		=
		\frac{1}{\pi}
		\exp\prt*{-\prt{v_1^2+v_2^2}}
		\prt*{v_1^2+v_2^2},
	\end{align}
	and the asymptotic steady state is the Maxwellian
	\begin{align}
		f_\infty(v_1,v_2)
		=
		\frac{1}{2\pi}
		\exp\prt*{-\frac{v_1^2+v_2^2}{2}},
	\end{align}
	for every exponent $\gamma$. However, the choice of exponent determines the evolution in time of $f$. We will study the Maxwellian ($\gamma=0$) and Coulombian ($\gamma=-d$) cases numerically.
}

\revisionOne{In the Maxwellian case, an exact solution exists (a derivation can be found in the Appendix of \cite{CHW2020}), which could also be use for validation. However, we prefer to compare both the Maxwellian and Coulombian cases on an equal footing; thus we shall study the numerical convergence of the method in both cases via relative errors.}

We consider the domain $v\in\prt*{-4,4}^2$, and solve the Landau equation with datum $f_0=f(0,v)$ for $t\in\prt*{0,15}$, initialising the particles with a stratified sampling. The collision strength is $C=2^{-4}$. We choose $\Nv\in\set{8,16,32,64}$ (note $N_{v_1}=N_{v_2}=\Nv$), $\Nc\in\set{1,2,4,8}$, $\Dt\in\set{100^{-1},200^{-1},400^{-1}}$. We solve the problems with and without random batching (respectively, $R=1$ and $R=16$).

\begin{figure}
	\centering
	\includegraphics[]{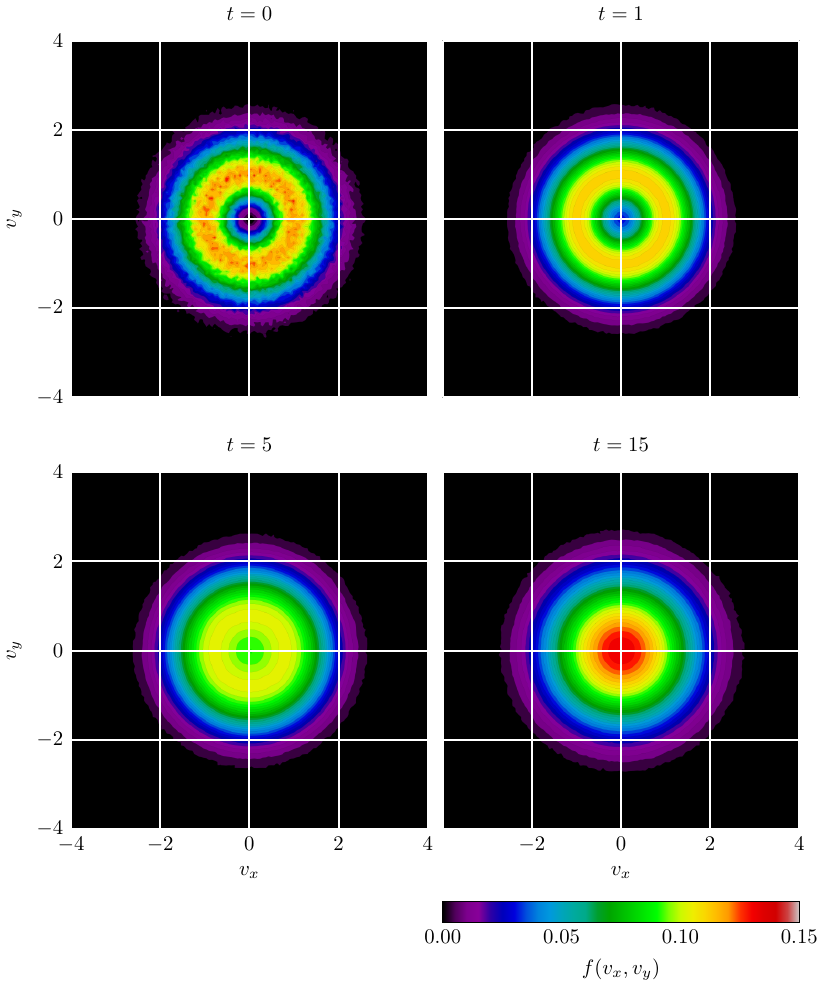}
	\caption{
	Typical solution in the Maxwellian case of the validation test of \cref{sec:BKW}.
	$C=2^{-4}$, $t\in\prt*{0,15}$.
	$\Nv=64$, $\Nc=8$, $\Dt=10^{-2}$.
	$R=16$.
	}
	\label{fig:validation}
\end{figure}

\Cref{fig:validation} shows a typical numerical solution. \Cref{fig:validation_order_b1} shows the relative $\Ltwo$ error of the solutions, defined for each value of $\Nv$ and its corresponding solution $f_{\Nv}$ as
\begin{align}
	\text{Error}(\Nv) \coloneqq \frac{\pnorm{2}{f_{\Nv} - f_{\Nv/2}}}{\pnorm{2}{f_{\Nv}}}.
\end{align}
\revisionOne{
Interestingly, Coulomb performs better than Maxwell when the number of particles per cell is very low ($\Nc=1$); in the base case, $\Dt=100^{-1}$, the Maxwellian case appears not to converge; this can be resolved by reducing $\Dt$. However, for $\Nc\geq 4$, both cases display second order convergence with respect to $\Nv$, and the error is lower for the Maxwellian case. The use of random batches ($R=16$) barely affects these results.
}
 
\subsubsection{Landau Damping (1D-2V)}\label{sec:LandauDamping}

We now validate the inhomogeneous scheme by studying the Landau damping \cite{Landau1936,Villani2013} of an electric wave. We adapt a test from \cite{MPZ2023}, suitably extended to the 1D-2V setting (similar tests have been used, for instance, in \cite{CF2004,CGM2013,ZG2017}). We verify that our numerical solution behaves consistently with respect to the known physical properties. Furthermore, we study the error in the entropy \revisionOne{increment} equality. \revisionOne{This is a necessary check because, as reported in \cite{TCT2022}, collisionless PIC simulations can in fact cause the entropy to increase. We establish here that any error due to the discretisation of the transport is several orders of magnitude smaller than the typical entropy values, and therefore the effects in our simulations are solely due to the presence of the collision operator.
}

The test considers a distribution in Maxwellian equilibrium, with a small spatial perturbation:
\begin{align}
	f_0(x,v_1,v_2)
	=
	\frac{1+\alpha\cos\prt*{kx}}{2\pi}
	\exp\prt*{-\frac{v_1^2+v_2^2}{2}},
\end{align}
with $k=2^{-1}$, and $\alpha=10^{-1}$. We will study the Coulombian case ($\gamma=-d$). Linear theory predicts that the $\Ltwo$ norm of the electric field will oscillate, but the amplitude of the oscillation will decay exponentially with rate
\begin{align}
	\gamma_l
	=
	-\frac{1}{k^3}\sqrt{\frac{\pi}{8}} \exp\prt*{-\frac{1}{2k^2}-\frac{3}{2}}
\end{align}
in the collisionless case. In the presence of sufficiently weak collisions, the decay is accelerated \revisionOne{by a linear correction to $\gamma_l+\revisionOne{C\gamma_{l,c}}$} \cite{DB1994,Chen2016}, where
\begin{align}
	\revisionOne{
		\gamma_{l,c}
		=
		-\sqrt{\frac{2}{9\pi}}
	}
\end{align}
This is, however, not the case for stronger collisions. \revisionOne{In the hydrodynamic limit ($C\rightarrow\infty$), in the absence of a magnetic field, the limiting behaviour of system \eqref{eq:continuity} is described by the Euler-Poisson equations, where the corresponding initial condition will result in a standing electric wave. Therefore, the actual decay rate of the electric field can be made arbitrarily slow for sufficiently strong collisions.
}

We consider the domain $x\in\prt*{0,2\pi/k}$, $v\in\prt*{-4,4}^2$, and solve the \acl{VAL} equation for $t\in\prt*{0,10}$, initialising the particles with a combined sampling (stratified in position, Gaussian in velocity). The collision strength is $C\in\set{0,0.01,0.1,1}$. We choose $\Nx=128$, $\Nv=32$ (note, once again, $N_{v_1}=N_{v_2}=\Nv$), $\Nc=8$, $\Dt=50^{-1}$, and $R=32$. Each simulation employs a total of $1\,048\,576$ particles.

\begin{figure}
	\centering
	\includegraphics[]{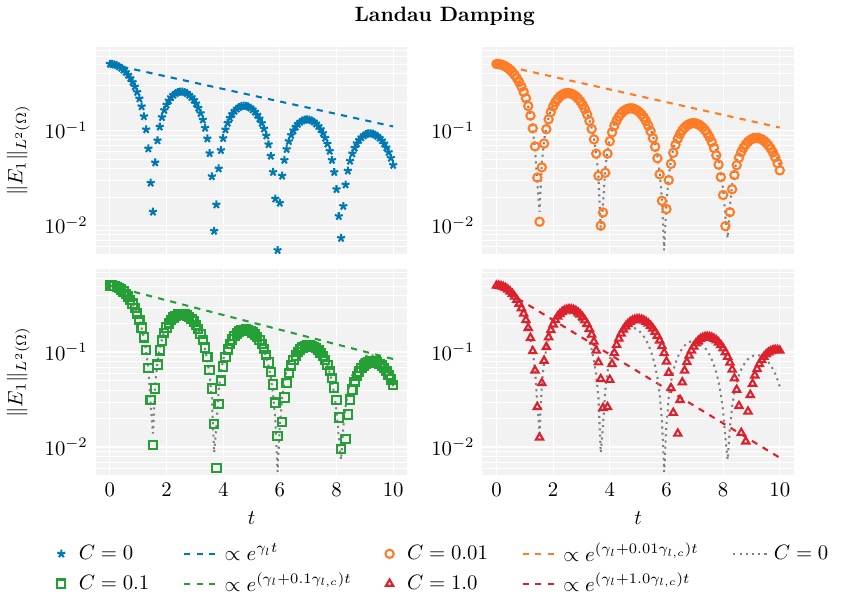}
	\caption{
	Numerical Landau damping in the validation test of \cref{sec:LandauDamping}.
	$t\in\prt*{0,10}$.
	$\Nx=128$, $\Nv=32$, $\Nc=8$, $\Dt=50^{-1}$.
	$R=32$.
	Total of $1\,048\,576$ particles.
	\revisionOne{Constants $\gamma_l$ and $\gamma_{l,c}$ given in \cref{sec:LandauDamping}.}
	}
	\label{fig:ld_decay}
\end{figure}

\Cref{fig:ld_decay} shows the decay of the first component of the electric field $(E_1)$ in the $\Ltwo$ norm. The collisionless case behaves as expected. The very weakly collisional case ($C=0.01$) is almost indistinguishable from collisionless; so much so, that it obscures the \revisionOne{reference} $C=0$ plot. The weakly collisional case ($C=0.1$) is within the purview of the linear analysis, and it shows an accelerated decay consistent with the theoretical rate. Finally, the strongly collisional case ($C=1.0$) breaks the linear trend and displays a decay rate slower than the collisionless one, very far from the linear prediction, \revisionOne{which is expected}.

\begin{figure}
	\centering
	\includegraphics[]{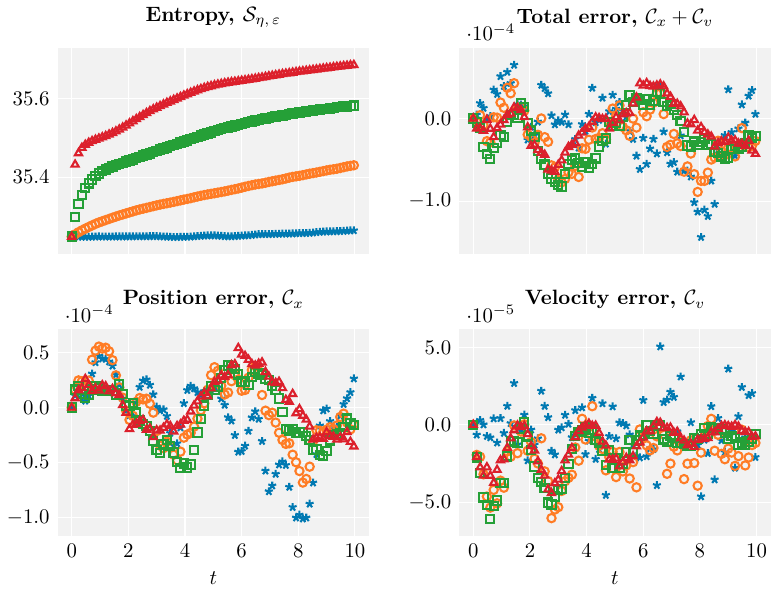}
	\caption{
	Entropy and entropy transport error in the Landau damping validation test of \cref{sec:LandauDamping}.
	$t\in\prt*{0,10}$.
	$\Nx=128$, $\Nv=32$, $\Nc=8$, $\Dt=50^{-1}$.
	$R=32$.
	Total of $1\,048\,576$ particles.
	}
	\label{fig:ld_energy}
\end{figure}

\revisionTwo{\Cref{fig:ld_energy} shows the} \revisionOne{increment} of the entropy in the collisional cases, as well as the entropy conservation errors discussed in \cref{sec:dissipation_entropy}. Compared to the upcoming experiments, the distribution $f$ in this test remains close to Maxwellian equilibrium as it evolves in time. Therefore, we expect the entropy \revisionOne{increment} effects to be relatively small. Nevertheless, we observe that the entropy transport error is at least three orders of magnitude smaller than the typical entropy \revisionOne{increment}; \revisionOne{in this scenario, the evolution of the entropy is driven solely by the action of the collision operator.}
 
\subsection{Study of Collisional Effects}\label{sec:collisions}

\subsubsection{Two-stream Instability (1D-2V)}\label{sec:TSI}

We turn to study the effects of the collisional effects in more interesting problems. Here we study the two-stream instability, a known interaction between two beams of electrons which causes a vortex to appear in phase space. We observe the smoothing of the distribution due to collisional effects, as well an improvement in the conservation of energy.

The test, adapted from \cite{MPZ2023}, considers two Maxwellian travelling beams, with a very small spatial perturbation:
\begin{align}
	f_0(x,v_1,v_2)
	=
	\frac{1+\alpha\cos\prt*{kx}}{2\pi}
	\brk*{
		\exp\prt*{-\frac{\prt*{v_1-c}^2}{2}}
		+
		\exp\prt*{-\frac{\prt*{v_1+c}^2}{2}}
	}
	\exp\prt*{-\frac{v_2^2}{2}},
\end{align}
with $c=2.4$, $k=5^{-1}$, and $\alpha=200^{-1}$. The electric field is initialised self-consistently. We study the Coulombian case ($\gamma=-2$).

We consider the domain $x\in\prt*{0,2\pi/k}$, $v\in\prt*{-6,6}^2$, and solve \acl{VAL} for $t\in\prt*{0,50}$, initialising the particles with a combined sampling (stratified in position, Gaussian in velocity). The collision strength is $C\in\set{0,0.01,0.02,0.04,0.08}$. We choose $\Nx=256$, $\Nvx=32$, $\Nvy=4$, $\Nc=16$, $\Dt=20^{-1}$, and $R=32$. Each simulation employs a total of $524\,288$ particles.

\begin{figure}
	\centering
	\includegraphics[]{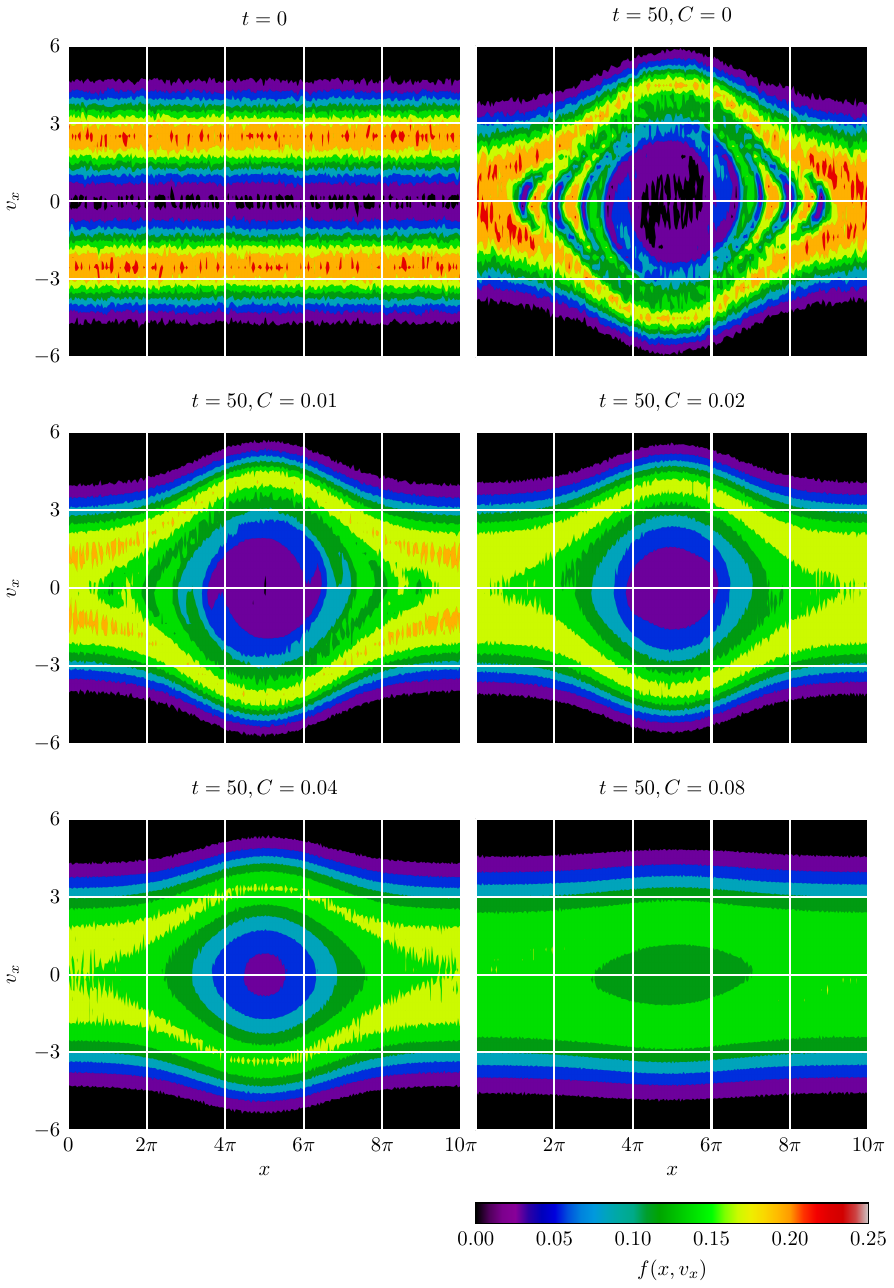}
	\caption{
	Final vortex in the two-stream instability test of \cref{sec:TSI} \revisionTwo{(marginals of $\ftN$)}.
	$t\in\prt*{0,50}$.
	$\Nx=256$, $N_{v_1}=32$, $N_{v_2}=4$, $\Nc=16$, $\Dt=20^{-1}$.
	$R=32$.
	Total of $524\,288$ particles.
	See \cite{BCH2024Web,BCH2024Fig} for animations.
	}
	\label{fig:TSIComparison}
\end{figure}

\Cref{fig:TSIComparedEvolution} shows the vortex formation in the collisionless setting, and the smoothing of the vortex as the effects of collision are strengthened; the action of the Landau operator as a non-linear, non-local diffusion operator is evident here. For completeness, \Cref{fig:TSIComparison} shows a comparison of the final state of the vortex for each collision strength. As expected, increasing $C$ interpolates from the collisionless dynamics to (nearly) a Maxwellian equilibrium.

\begin{figure}
	\centering
	\includegraphics{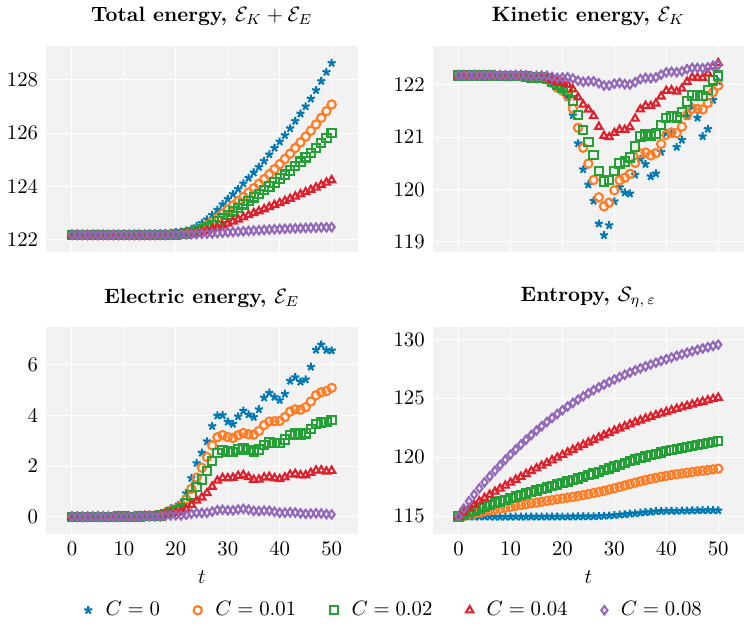}
	\caption{
	Energy and entropy in the two-stream instability test of \cref{sec:TSI}.
	$t\in\prt*{0,50}$.
	$\Nx=256$, $N_{v_1}=32$, $N_{v_2}=4$, $\Nc=16$, $\Dt=20^{-1}$.
	$R=32$.
	Total of $524\,288$ particles.
	See \cite{BCH2024Web,BCH2024Fig} for animations.
	}
	\label{fig:TSIEnergy}
\end{figure}

\Cref{fig:TSIEnergy} shows the evolution of the total energy, kinetic energy, electric energy, and entropy of the problem. Around time $20$, we see an exponential conversion of kinetic energy to electric energy: this corresponds to the initial appearance of the vortex. The conversion saturates around time $30$, but both energies continue to increase, thus the total energy is not exactly conserved. Due to the structure-preserving properties of our discretisation, stronger collisions lead to better conservation properties, not worse. The entropy \revisionOne{increases} throughout. \revisionOne{In the collisionless case, the entropy is essentially constant, until the point where the vortex appears, where some variation can be seen. This could be due to larger entropy conservation errors $\mathcal{C}_x$ and $\mathcal{C}_v$, or the PIC entropy increase effects described in \cite{TCT2022}.}

\begin{figure}
	\centering
	\includegraphics{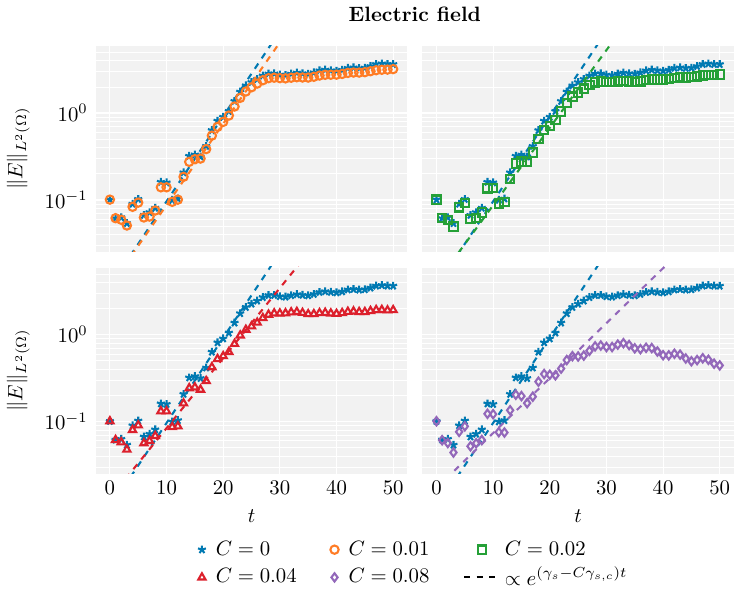}
	\caption{
	Exponential growth of the electric field in the two-stream instability test of \cref{sec:TSI}.
	$t\in\prt*{0,50}$.
	$\Nx=256$, $N_{v_1}=32$, $N_{v_2}=4$, $\Nc=16$, $\Dt=20^{-1}$.
	$R=32$.
	Total of $524\,288$ particles.
	See \cite{BCH2024Web,BCH2024Fig} for animations.
	\revisionOne{Constants $\gamma_s$ and $\gamma_{s,c}$ given in \cref{sec:TSI}.}
	}
	\label{fig:TSI}
\end{figure}

\revisionTwo{\Cref{fig:TSI} shows the growth of the $L^2$ norm of the electric field in time. In the collisionless case, the growth rate in the exponential phase is $\gamma_s = 0.2258$ \cite{Chen2016,LX2017,XF2021,MPZ2023}. It has recently been suggested \cite{ZB2023} that this growth rate should be insensitive to collisional effects; however, other sources \cite{SKV2016} expect the rate to slow down linearly for weak collisions to $\gamma_s-C\gamma_{s,c}$, for some positive constant $\gamma_{s,c}$. Our results are roughly consistent with $\gamma_{s,c}\simeq 1$.
}

\begin{figure}
	\centering
	\includegraphics[]{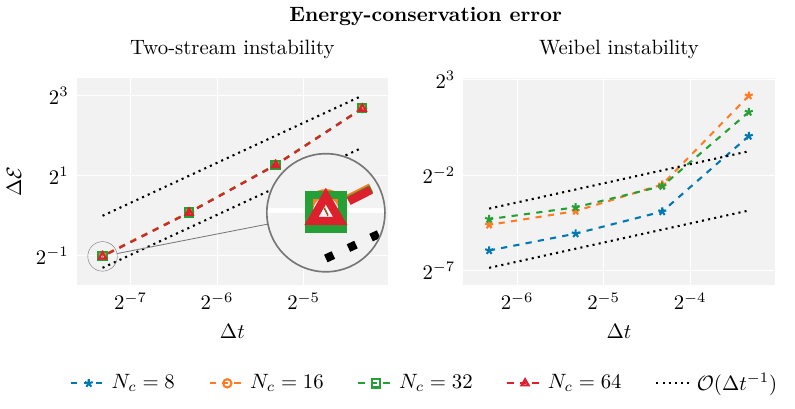}
	\caption{
		Collisionless conservation-of-energy error on two-stream instability (\cref{sec:TSI}) and the Weibel instability (\cref{sec:WI}) as a function of $\Dt$. Starting from the original simulation parameters, $\Nc$ is doubled twice, and $\Dt$ is halved thrice; other parameters remain unchanged.
	}
	\label{fig:EnergyValidation}
\end{figure}

\revisionTwo{\Cref{fig:EnergyValidation} shows the energy-conservation error for the collisionless experiment, defined simply as the absolute difference between the final and initial (discrete) energy. The collisionless simulation is repeated both with a larger number of particles (doubling $\Nc$ twice) and with a smaller time step (halving $\Dt$ thrice) in order to study whether the conservation error decreases. In this test, the error is essentially insensitive to the number of particles; however, it decreases linearly with $\Dt$. This is expected, in view of the discussion of \cref{sec:global_properties_conservation} and the fact that we employ an explicit Euler integrator. As already remarked in the Introduction and \cref{sec:global_properties_conservation}, better time integrations are available, and would lead to improved conservation properties, but their use in combination with our collision operator is too costly at present. Nevertheless, we conclude that the use of more sophisticated integrators or a finer time step would lead to much better conservation in our numerics.
} 
\subsubsection{Weibel Instability (1D-2V)}\label{sec:WI}

To conclude, we study the Weibel instability, a problem which includes the full electromagnetic effects. Once again we observe the smoothing of the distribution due to collisional effects, as well as improved conservation properties.

This test, borrowed from \cite{CCZ2014b}, considers two Maxwellian beams
\begin{align}
	f_0(x,v_1,v_2)
	=
	\revisionTwo{
		\frac{1}{\pi\beta}
		\exp\prt*{-\frac{v_1^2}{\beta}}
		\brk*{
			\exp\prt*{-\frac{\prt*{v_2-c}^2}{\beta}}
			+
			\exp\prt*{-\frac{\prt*{v_2+c}^2}{\beta}}
		},}
\end{align}
with $c=0.3$, and $\beta=10^{-2}$. The electric field is initialised self-consistently (to zero), and the initial magnetic field is
\begin{align}
	B_3(0,x) = \alpha \sin\prt*{kx},
\end{align}
where $k=5^{-1}$, and $\alpha=10^{-3}$. We study the Coulombian case ($\gamma=-2$).

We consider the domain $x\in\prt*{0,2\pi/k}$, $v\in\prt*{-6,6}^2$, and solve \acl{VML} for $t\in\prt*{0,125}$, initialising the particles with a combined sampling (stratified in position, Gaussian in velocity). The collision strength is $C\in\set{0,0.0001,0.0002,0.0004,0.0008}$. We choose $\Nx=32$, $\Nvx=64$, $\Nvy=64$, $\Nc=8$, $\Dt=10^{-1}$, and $R=64$. Each simulation employs a total of $1\,048\,576$ particles.

\begin{figure}
	\centering
	\includegraphics[]{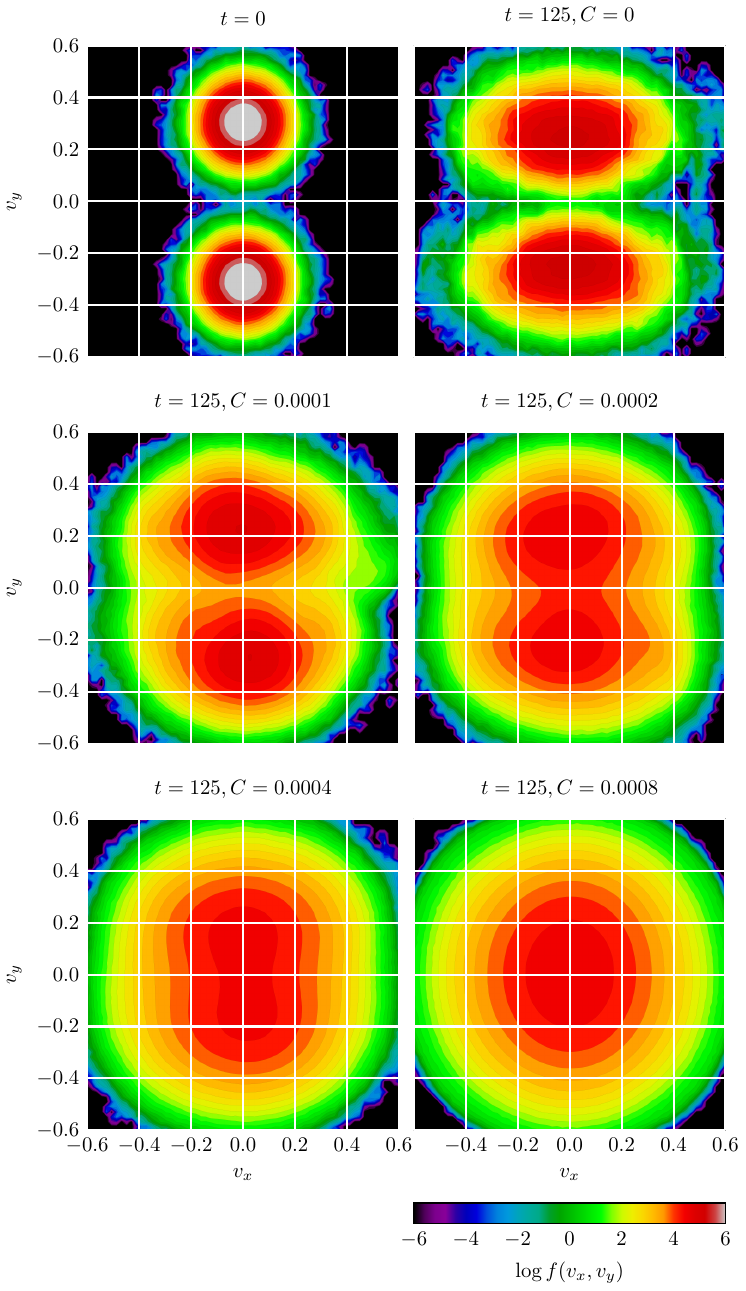}
	\caption{
	Final beam collapse in the Weibel instability test of \cref{sec:WI} \revisionTwo{(marginals of $\ftN$)}.
	$t\in\prt*{0,125}$.
	$\Nx=32$, $\Nv=64$, $\Nc=8$, $\Dt=10^{-1}$.
	$R=64$.
	Total of $1\,048\,576$ particles.
	See \cite{BCH2024Web,BCH2024Fig} for animations.
	}
	\label{fig:WIComparison}
\end{figure}

\revisionTwo{\Cref{fig:WIComparedEvolution}} shows the action of the instability. Without collisional effects, the initially narrow, fast beams suddenly become wider and slow, coinciding with the exponential growth of the magnetic field. The new beams are stable, though they appear noisy in the figure due to the nature of the particle approximation. The collisional cases display similar qualitative behaviour at the beginning, but eventually the two beams begin to merge. Once again, the diffusive nature of the Landau operator is evident. For completeness, \Cref{fig:WIComparison} shows a comparison of the final state of the beams for each collision strength. As before, stronger collisions drive the distribution towards a Maxwellian.

\begin{figure}
	\centering
	\includegraphics{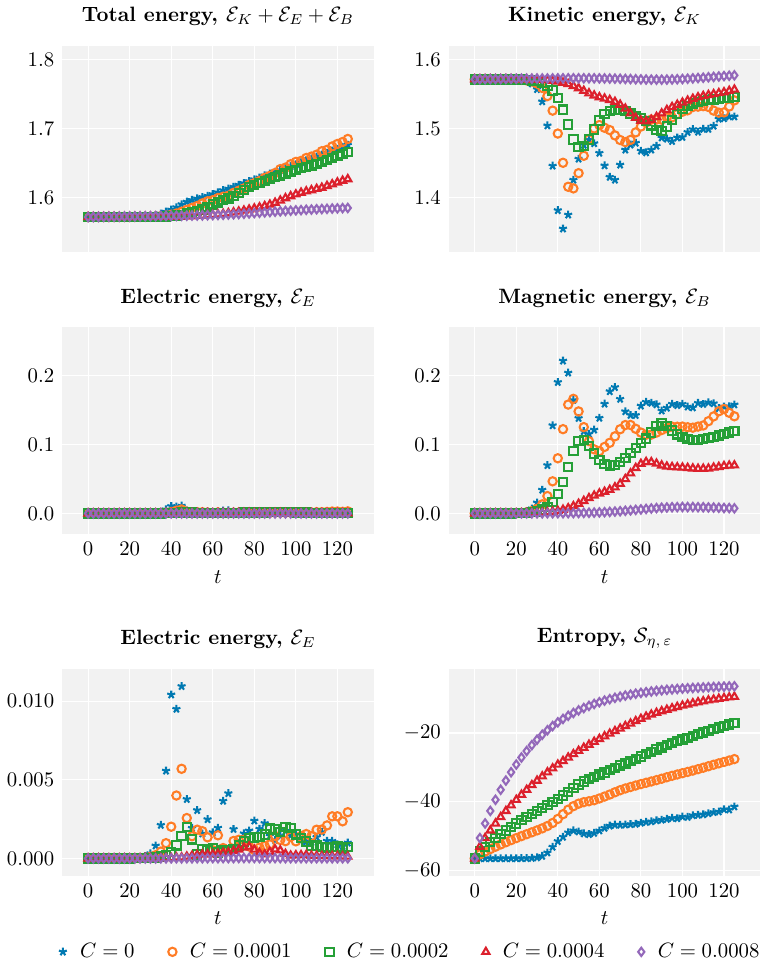}
	\caption{
	Energy and entropy in the Weibel instability test of \cref{sec:WI}.
	$t\in\prt*{0,125}$.
	$\Nx=32$, $\Nv=64$, $\Nc=8$, $\Dt=10^{-1}$.
	$R=64$.
	Total of $1\,048\,576$ particles.
	See \cite{BCH2024Web,BCH2024Fig} for animations.
	}
	\label{fig:WIEnergy}
\end{figure}

\Cref{fig:WIEnergy} shows the evolution of the total energy, kinetic energy, electric energy, magnetic energy, and entropy of the problem. Around time $30$, we see an exponential conversion of kinetic energy to magnetic energy. The conversion saturates around time $40$, when the magnetic energy begins to oscillate around a constant value, and appears convergent in time. The electric energy behaves similarly, but its magnitude is 20 times smaller, establishing this as a mainly magnetic effect. Unlike the two-stream instability, the lack of conservation here is mainly visible in the kinetic energy, which grows steadily. Once again, stronger collisions lead to better conservation properties, not worse. \revisionOne{Just as in the two-stream instability case, the entropy conservation error due to the transport is no longer negligible once the instability develops, leading to variations in the entropy in the collisionless case. Nevertheless, the addition of collisional terms still play an important role in the evolution of the entropy, which increases monotonically when $C$ is large enough.}

\revisionTwo{Just as in the previous section, \Cref{fig:EnergyValidation} shows the energy-conservation error for the collisionless experiment. Again, the collisionless simulation is repeated both with a larger number of particles and with a smaller time step. In this test, the error is not insensitive to the number of particles, but it nevertheless decreases linearly with $\Dt$. Once again, we conclude that the use of more advanced time integrators or a smaller $\Dt$ would greatly improve the conservation errors across our experiments.
}  
\section{Conclusion \& Outlook}\label{sec:conclusion}

In this work, we have introduced a generalisation of the \acf{PIC} method able to simulate the Landau collisions in a spatially inhomogeneous plasma. We have derived the method from the gradient-flow structure of the Landau equation, preserving a variational structure, and retaining all conservation properties \revisionTwo{for the collisional update}. We have validated the numerical method, and presented collisional simulations of phenomena such as the Landau damping, the two-stream instability, and the Weibel instability, which demonstrate its ability to simulate a collisional plasma.

Several extensions of the work could be considered. A clear future task is to perform 3D-3V simulations of a collisional plasma with full electromagnetic effects using \ac{CPIC}; this remains a tough computational challenge, and will likely involve the implementation of \ac{CPIC} within a mature plasma simulation ecosystem, such as NESO \cite{TAA2023}. It would also be of interest to study whether the techniques employed in this work could be applied to the gyrokinetic Landau operator \cite{KHC2016}, as a way to reduce the overall computational cost. Furthermore, a natural objective is the extension of the method to a two species setting, as \cite{CHV2023} did for the homogeneous scheme of \cite{CHW2020}, though the large ion-electron mass ratio will likely lead to challenging restrictions on the time integrator; a similar challenge was addressed in \cite{BR2022} for a different class of schemes and collision operators. Following \cite{BCM2023}, stochastic Galerkin expansions could be used to perform uncertainty quantification at the inhomogeneous level, as was done in \cite{MPZ2023} for the Vlasov-Poisson-BGK equations. The use of \ac{CPIC} as a source of synthetic data to train a surrogate model is not out of the question; \cite{MCD2021} performs a similar task at the gyrokinetic level with data from XGC1 \cite{CK2008,KCD2009}.

\section*{Acknowledgements}
RB and JAC were supported by the Advanced Grant Nonlocal-CPD (Nonlocal PDEs for Complex Particle Dynamics: Phase Transitions, Patterns and Synchronization) of the European Research Council Executive Agency (ERC) under the European Union’s Horizon 2020 research and innovation programme (grant agreement No.~883363) and by the EPSRC grant EP/T022132/1 ``Spectral element methods for fractional differential equations, with applications in applied analysis and medical imaging''. JH was partially supported by AFOSR grant FA9550-21-1-0358 and DOE grant DE-SC0023164. 
\section*{Appendix: Dimensionless Vlasov-Maxwell-Landau equations for the Coulomb case in dimension three}

For the sake of completeness, we describe here the non-dimensionalisation that leads to \eqref{eq:continuity}, which roughly follows \cite{Degond2007}. Throughout, we assume we are in the Coulomb setting, $\gamma=-\dimv=-3$.

To begin, we assume the typical density is $n_0$ and the typical temperature is $T_0$. The \textit{thermal speed} is defined as
\begin{align}
	v_0 \coloneqq \sqrt{\frac{k_B T_0}{m}},
\end{align}
and the typical distribution magnitude is defined as $f_0 = n_0 v_0^{-3}$ for dimensional consistency, which results in the scaling $f = f_0 \bar{f}$.

The velocity coordinate will be scaled by $v = v_0 \bar{v}$. To define scales for time and space, we make use of the \textit{Debye length},
\begin{align}
	\lambda_D = \sqrt{\frac{\varepsilon_0 k_B T_0}{n_0 q^2}},
\end{align}
and the \textit{plasma frequency},
\begin{align}
	\omega_p \coloneqq v_0 / \lambda_D = \sqrt{\frac{n_0q^2}{m\varepsilon_0}}.
\end{align}
We can now define $t_0 = \omega_p^{-1}$ and $x_0 = \lambda_D$, and scale $t = t_0 \bar{t}$ and $x = x_0 \bar{x}$.

The typical force has magnitude $F_0\coloneqq k_B T_0 x_0^{-1}$; the fields will therefore be scaled by $E = E_0 \bar{E}$ and $B = B_0 \bar{B}$, with scales $E_0 = F_0q^{-1}$ and $B_0 = E_0 v_0^{-1}$.

We assume the Coulomb collisional cross-section normalises with a factor
\begin{align}
	\sigma_0 = \frac{1}{8} \abs{ \log\delta } \frac{ q^4 }{m^2\pi\varepsilon_0^2 v_0^4}= \frac{\mathcal{C}_{-3}}{v_0^4},
\end{align}
see \cite{Degond2007} for details. We define the \textit{collision frequency} $\nu_0 \coloneqq \sigma_0 n_0 v_0$, and scale the collision
operator by $Q = Q_0 \bar{Q}$, where $Q_0\coloneqq \nu_0 f_0$.

We first scale Maxwell's equations. Amp\`{e}re's equation $\varepsilon_0 \mu_0 \pt E = \curlx B - \mu_0 J$ becomes $\partial_{\bar{t}} \bar{E} = \frac{c^2}{v_0^2}\grad_{\bar{x}} \times \bar{B} - \bar{J}$ in the new scale, where $J = J_0 \bar{J}$ and $J_0 = q v_0 n_0$ for consistency. We make here the non-relativistic assumption $c=v_0$ to arrive at $\partial_{\bar{t}} \bar{E} = \grad_{\bar{x}} \times \bar{B} - \bar{J}$. Poisson's equation $\varepsilon_0 \divx E = \rho + \rho\ion$ becomes $\grad_{\bar{x}} \cdot \bar{E} = \bar{\rho} - \bar{\rho}\ion$, where $\rho = \rho_0 \bar{\rho}$, $\rho_0 = q n_0$, and $\rho\ion = - \rho_0 \bar{\rho}\ion $. Faraday's and Gauss' equations remain unchanged in the new variables.

We now scale the Vlasov-Landau equation, $\pt f + v \cdot \gradx f + \frac{q}{m} \prt*{E + v\times B} \cdot \gradv f = Q[f,f]$. Under our assumptions, the equation becomes
\begin{align}
	\partial_{\bar{t}} \bar{f} + \bar{v} \cdot \grad_{\bar{x}} \bar{f} + \prt*{\bar{E} + \bar{v}\times \bar{B}} \cdot \grad_{\bar{v}} \bar{f} = \bar{Q}[\bar{f},\bar{f}],
\end{align}
where
\begin{align}
	\bar{Q}[\bar{f},\bar{f}] = \grad_{\bar{v}} \cdot \int_{\R^3} A(\bar{v}-\bar{v}_*) \brk*{ \bar{f}(\bar{v}_*) \grad_{\bar{v}} \bar{f}(\bar{v}) - \bar{f}(\bar{v}) \grad_{\bar{v}_*} \bar{f}(\bar{v}_*) } \dd \bar{v}_*.
\end{align}
The \textit{collisional cross-section} $A$ is a symmetric and semi-positive-definite matrix given by
\begin{align}
	A(z) = C \abs{z}^{-1} \Pi(z),
	\quad \Pi(z) = \prt*{
		I_{3} - \frac{z\otimes z}{\abs{z}^2}
	},
\end{align}
where $C=\nu_0 \omega_p^{-1}$ is the \textit{dimensionless collision strength}, obtained as the quotient of the collisional frequency and the plasma frequency.

 \begin{figure}
	\centering
	\includegraphics{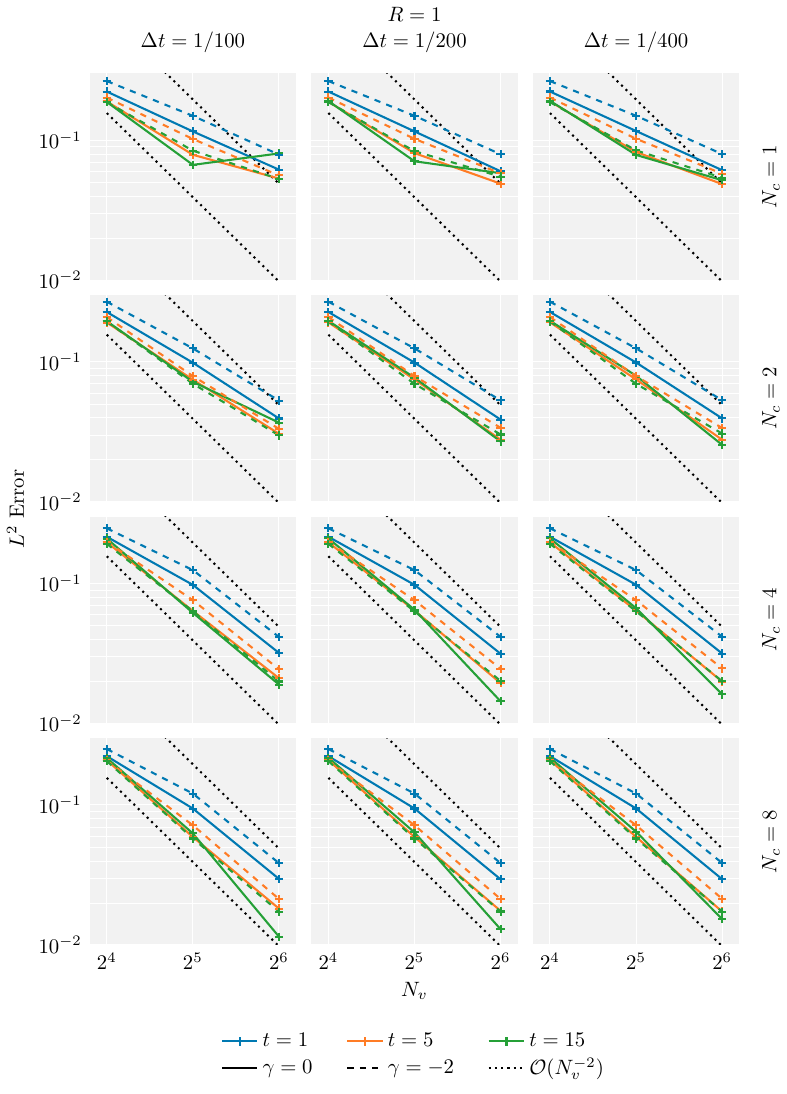}
	\caption{
	Order of convergence for the validation test of \cref{sec:BKW}. Relative $\Ltwo$ errors of the regularised solutions $\ftN$.
	$C=2^{-4}$, $t\in\prt*{0,15}$.
	}
	\label{fig:validation_order_b1}
\end{figure}

\begin{figure}
	\ContinuedFloat
	\captionsetup{list=off,format=cont}
	\centering
	\includegraphics{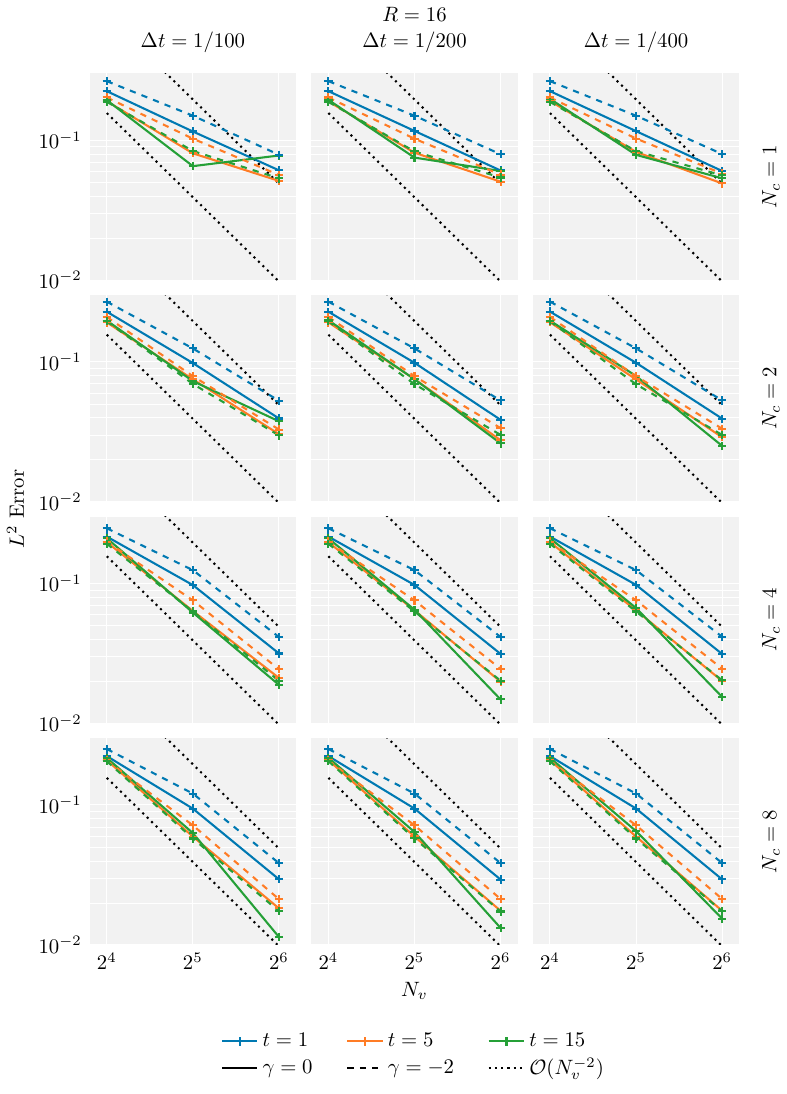}
	\caption{
	Order of convergence for the validation test of \cref{sec:BKW}. Relative $\Ltwo$ errors of the regularised solutions $\ftN$.
	$C=2^{-4}$, $t\in\prt*{0,15}$.
	}
	\label{fig:validation_order_b16}
\end{figure} 
  
\begin{figure}
	\centering
	\includegraphics[]{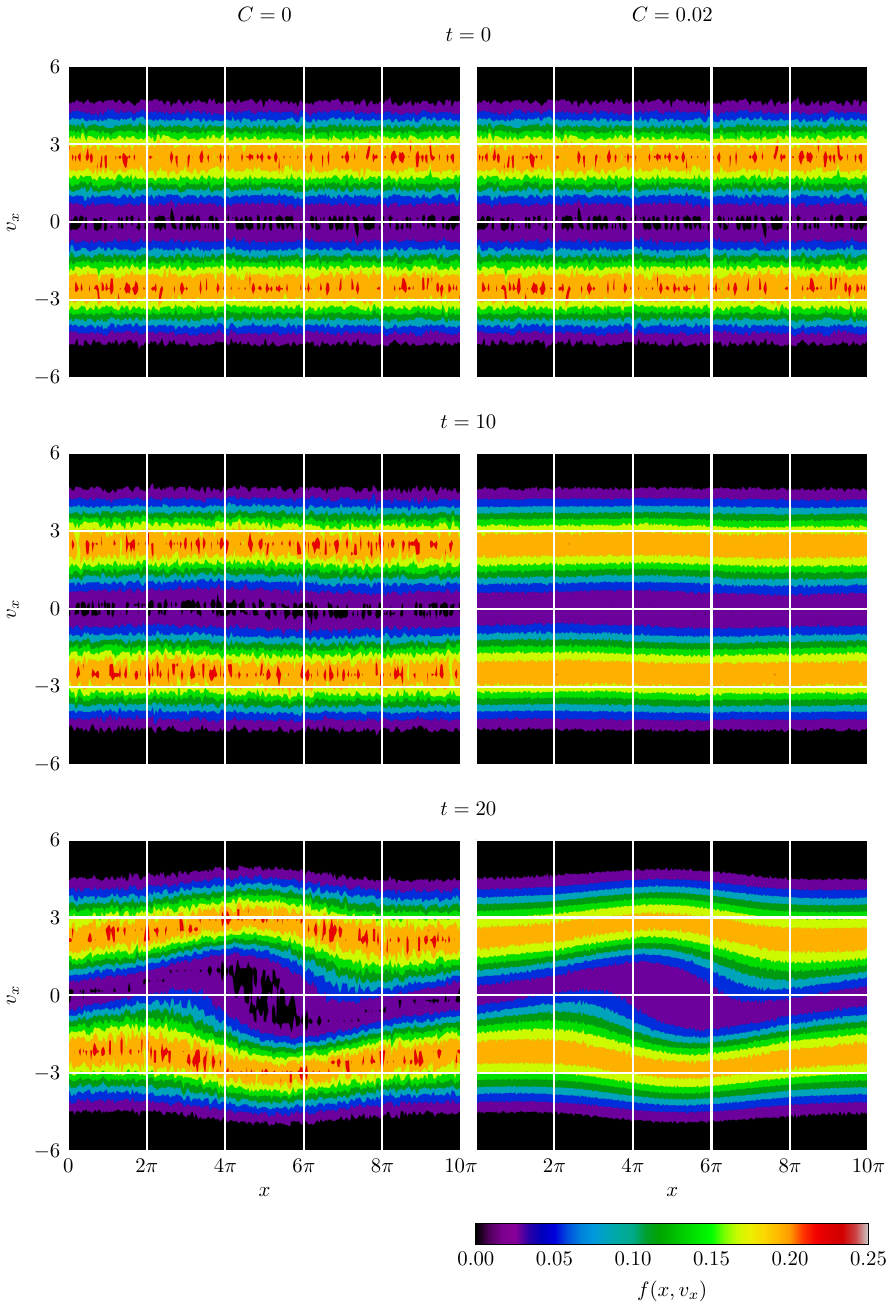}
	\caption{
	Vortex formation in the two-stream instability test of \cref{sec:TSI} \revisionTwo{(marginals of $\ftN$)}.
	$t\in\prt*{0,50}$.
	$\Nx=256$, $N_{v_1}=32$, $N_{v_2}=4$, $\Nc=16$, $\Dt=20^{-1}$.
	$R=32$.
	Total of $524\,288$ particles.
	See \cite{BCH2024Web,BCH2024Fig} for animations.
	}
	\label{fig:TSIComparedEvolution}
\end{figure}

\begin{figure}
	\ContinuedFloat
	\captionsetup{list=off,format=cont}
	\centering
	\includegraphics[]{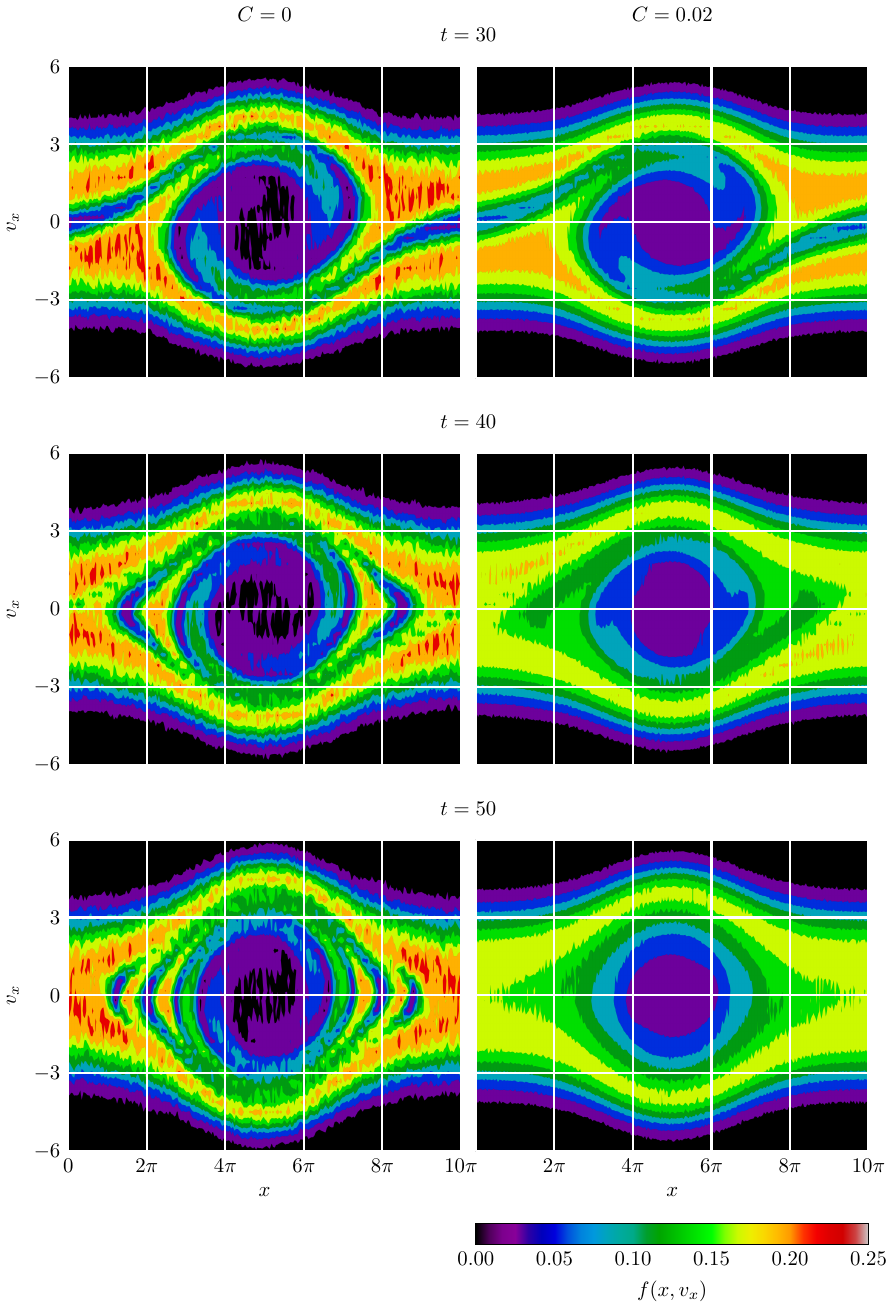}
	\caption{
	Vortex formation in the two-stream instability test of \cref{sec:TSI} \revisionTwo{(marginals of $\ftN$)}.
	$t\in\prt*{0,50}$.
	$\Nx=256$, $N_{v_1}=32$, $N_{v_2}=4$, $\Nc=16$, $\Dt=20^{-1}$.
	$R=32$.
	Total of $524\,288$ particles.
	See \cite{BCH2024Web,BCH2024Fig} for animations.
	}
\end{figure} 
 
\begin{figure}
	\centering
	\includegraphics[]{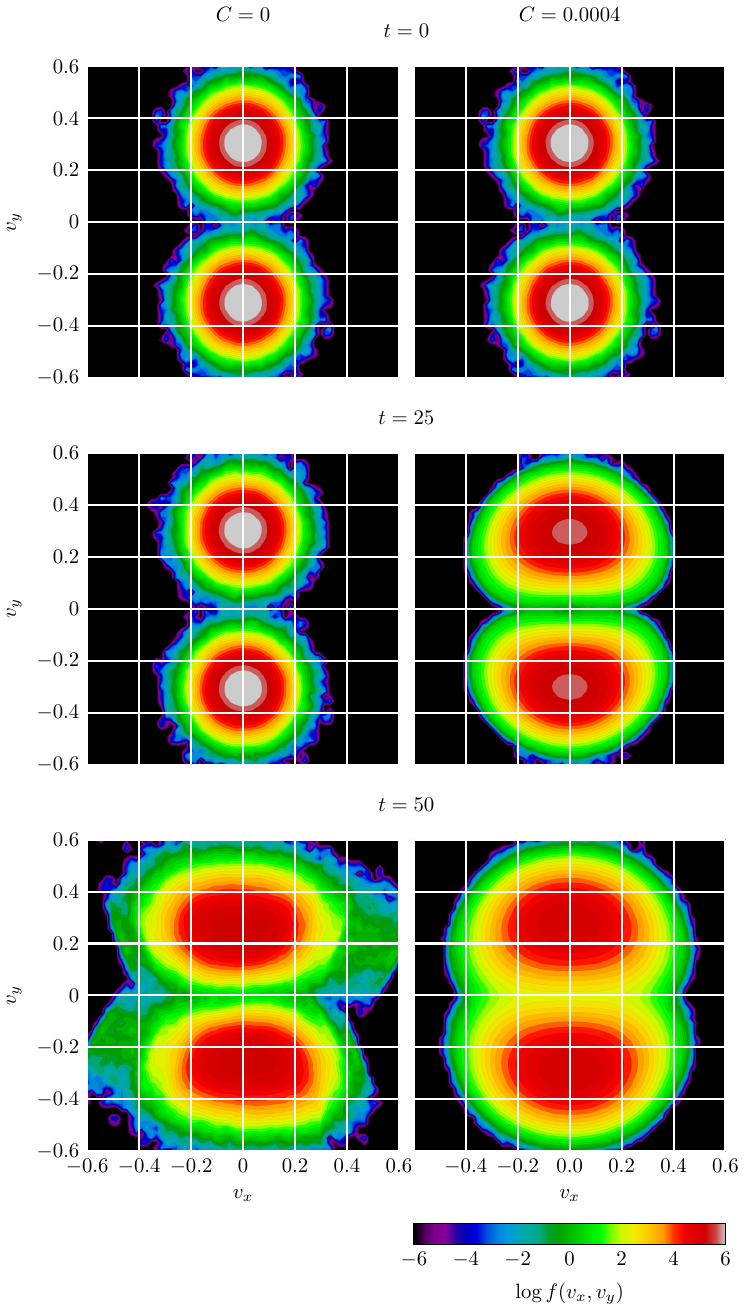}
	\caption{
	Electron beam collapse in the Weibel instability test of \cref{sec:WI} \revisionTwo{(marginals of $\ftN$)}.
	$t\in\prt*{0,125}$.
	$\Nx=32$, $\Nv=64$, $\Nc=8$, $\Dt=10^{-1}$.
	$R=64$.
	Total of $1\,048\,576$ particles.
	See \cite{BCH2024Web,BCH2024Fig} for animations.
	}
	\label{fig:WIComparedEvolution}
\end{figure}

\begin{figure}
	\ContinuedFloat
	\captionsetup{list=off,format=cont}
	\centering
	\includegraphics[]{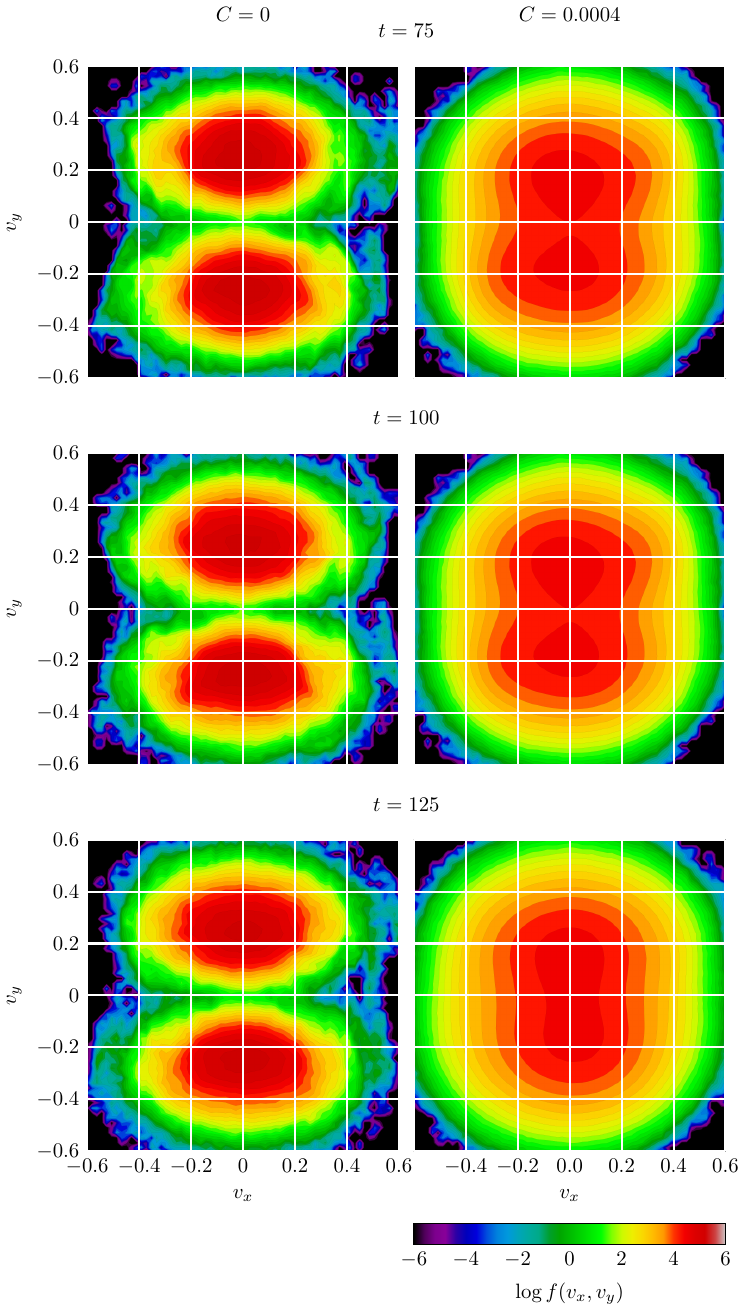}
	\caption{
	Electron beam collapse in the Weibel instability test of \cref{sec:WI} \revisionTwo{(marginals of $\ftN$)}.
	$t\in\prt*{0,125}$.
	$\Nx=32$, $\Nv=64$, $\Nc=8$, $\Dt=10^{-1}$.
	$R=64$.
	Total of $1\,048\,576$ particles.
	See \cite{BCH2024Web,BCH2024Fig} for animations.
	}
\end{figure} 

{
\small
\bibliographystyle{./abbrv_mod}
\bibliography{BailoCarrilloHu_CPIC}
\addcontentsline{toc}{section}{References}
}
 
\end{document}